\documentclass[aps,prb,reprint,groupedaddress]{revtex4-1}
\usepackage[english]{babel}
\usepackage{xcolor}
\usepackage{amsmath}
\usepackage{mathtools}
\usepackage{amssymb}
\usepackage{physics}
\usepackage[utf8]{inputenc}
\usepackage{lmodern}
\usepackage{graphicx}
\usepackage{grffile}
\usepackage{hyperref}
\usepackage[footnote=true]{snotez}
\usepackage{changes}

\newcommand{\gate}[1]{\texttt{#1}}

\definechangesauthor[name={Sebastian Zanker}]{SZ}
\definechangesauthor[name={Nicolas Vogt}]{NV}

\begin{document}
\title{Preparing symmetry broken ground states with variational quantum algorithms}
\author{Nicolas Vogt}
\email[]{Nicolas.Vogt@quantumsimulations.de}
\homepage[]{www.quantumsimulations.de}
\affiliation{HQS Quantum Simulations GmbH \\
Haid-und-Neu-Straße 7 \\
76131 Karlsruhe, Germany}
\author{Sebastian Zanker}
\affiliation{HQS Quantum Simulations GmbH \\
Haid-und-Neu-Straße 7 \\
76131 Karlsruhe, Germany}
\author{Jan-Michael Reiner}
\affiliation{HQS Quantum Simulations GmbH \\
Haid-und-Neu-Straße 7 \\
76131 Karlsruhe, Germany}
\author{Thomas Eckl}
\affiliation{Robert Bosch GmbH \\
Robert-Bosch-Campus 1\\
71272 Renningen, Germany}
\author{Anika Marusczyk}
\affiliation{Robert Bosch GmbH \\
Robert-Bosch-Campus 1\\
71272 Renningen, Germany}
\author{Michael Marthaler}
\affiliation{HQS Quantum Simulations GmbH \\
Haid-und-Neu-Straße 7 \\
76131 Karlsruhe, Germany}
\date{\today}
\begin{abstract}
  One of the most promising applications for near term quantum computers is the simulation of physical quantum systems, particularly many-electron systems in chemistry and condensed matter physics.
  In solid state physics, finding the correct symmetry broken ground state of an interacting electron
  system is one of the central challenges.
  The Variational Hamiltonian Ansatz (VHA), a variational hybrid quantum-classical algorithm especially suited for finding the ground state of a solid state system, will in general not prepare a broken symmetry state unless the initial state is chosen to exhibit the correct symmetry.
  In this work, we discuss three variations of the VHA designed to find the correct broken symmetry states close to a transition point between different orders.
  As a test case we use the two-dimensional Hubbard model where we break the symmetry explicitly by means of external fields coupling to the Hamiltonian and calculate the response to these fields.
  For the calculation we simulate a gate-based quantum computer and also consider the effects of dephasing noise on the algorithms.
  We find that two of the three algorithms are in good agreement with the exact solution for the considered parameter range.
  The third algorithm agrees with the exact solution only for a part of the parameter regime, but is more robust with respect to dephasing compared to the other two algorithms.

\end{abstract}
\maketitle
\section{Introduction}
The simulation of fermionic quantum systems is one of the most promising applications for near-term noisy intermediate scale quantum computers (NISQ).
An important use-case is computing the ground state and the ground state energy of an electron system, for example, the Hamiltonian of a molecule or a condensed matter system.

The currently most promising algorithms for near term quantum computers are the Variational Quantum Eigensolvers (VQE) \cite{peruzzo_variational_2014}.
These are hybrid quantum-classical algorithms combining a quantum computation of the energy expectation value of a parameterized trial state with a classical optimization.
Due to the combination of classical parameter optimization and quantum computations, the typical circuit width and depth of a VQE are significantly lower than for quantum phase estimation (QPE), another well-known algorithm to find the ground state of a quantum system \cite{wecker_solving_2015, nielsen_quantum_2011, Wecker_vha_15}.

Different types of VQE are mainly distinguished by the ansatz for the trial states.
The two most widely used variants are the unitary Coupled Cluster Single Double ansatz (uCCSD) and the Variational Hamiltonian Ansatz (VHA) \cite{Romero_ucc_19, Wecker_vha_15}.
In recent years, there has been continuous progress in the development of VQE methods including extensions to find excited states \cite{mcclean_theory_2016, mcclean_hybrid_2017,parrish_quantum_2019, nakanishi_subspace-search_2019}.
For an overview, we refer the reader to Ref.\onlinecite{sam_mcardle_quantum_2018}.

In this work, we focus on the Hamiltonian Ansatz for variational algorithms and its application to solid state systems with broken symmetries.
Understanding broken symmetries and the corresponding phase transitions in solid state systems is one of the central problems in condensed matter physics.
Being able to identify the correct order of the ground state of a large simulated system near a (quantum) critical phase transition with the help of a quantum computer could have an immediate impact on a wide field of research  ranging from high \(T_C\) superconductivity to magnetism.

The VHA is especially well-suited for condensed matter problems.
The Hamiltonian of a solid state system without disorder can often be mapped to model Hamiltonians that -- due to its lattice structure -- can be decomposed into a small number of partial Hamiltonians.
This leads to a small number of classical parameters in the Variational Hamiltonian Ansatz.
Whether or not the hybrid quantum classical algorithm can detect the correct order of a ground state depends on the available trial states in the Variational Hamiltonian Ansatz.
In this work, we discuss three variants of the VHA that ensure that the set of possible trial states includes states with the correct broken symmetries.

In section \ref{sec:var_algo}, we give a short overview of the VHA and the symmetry properties of trial states and introduce the three variants used in this work: VHA with post-selection (VHAPS), Variational Extended Hamiltonian Ansatz (VEHA) and Variational Mean Field Hamiltonian Ansatz (VMFHA).
In section \ref{sec:hubbard}, we introduce the two-dimensional Hubbard model as a test case to compare the three approaches.
The computationally feasible system sizes of 3 by 2 Hubbard sites are too small to observe proper phase transitions.
We therefore use a model with explicitly broken symmetries and study the dominant response of order-parameters to external fields.
In section \ref{sec:numeric}, we present the results of full simulations of the three algorithms as they would run on a quantum computer.
For these results, the entire gate-based trial state preparation including initialization and noise has been simulated on a classical computer.

\section{Variational algorithms}
\label{sec:var_algo}
\subsection{The standard VHA and broken symmetries}
The Variational Hamiltonian Ansatz \cite{Wecker_vha_15} (VHA) is a variational algorithm designed to find the ground state of a system described by a Hamiltonian $H$.
The key idea of variational algorithms is to apply a unitary operator $U(\vec\theta)$ to an initial state $\ket{\psi_0}$ that is easy to prepare, which gives a trial state as a function of the classical parameters \(\vec\theta\),
\begin{equation}
    \ket{\psi( \vec\theta )} = U(\vec\theta) \ket{\psi_0} \ .
\end{equation}
The energy of the trial state then also depends on these parameters,
\begin{align}
E(\vec\theta) = \bra{\psi(\vec\theta)}  H  \ket{\psi(\vec\theta)} \ ,
\end{align}
and has the ground state energy \(E_{\mathrm{gs}}\) of the Hamiltonian \(H\) as a lower bound,
\begin{equation}
  \label{eq:var_p}
  E_{\mathrm{gs}} \le E(\vec\theta)\,.
\end{equation}
Minimizing the energy yields an optimal parameter set \(\vec\theta_\mathrm{min}\) such that
\begin{equation}
E(\vec\theta_\mathrm{min}) = \min_{\vec\theta} E(\vec\theta) \ .
\end{equation}

In the ideal case, the trial states span the whole Hilbert space of the problem and the
classical optimizer finds the true global minimum of \(E(\vec\theta)\).
In that case, the trial state corresponding to the optimized parameters,
\begin{equation}
  \psi_{\mathrm{VHA}} = \psi(\vec\theta_\mathrm{min}) \ ,
\end{equation}
is the true ground state.
In reality both assumptions will not hold and \(\psi_{\mathrm{VHA}}\) is merely an approximation of the true ground state.

The choice of the unitary transformation in the Variational Hamiltonian Ansatz is inspired by the adiabatic evolution.
The unitary transformations correspond to time-evolutions under partial Hamiltonians of the full Hamiltonian \(H\).
We assume that the Hamiltonian can be separated into \(N\) partial Hamiltonians, \(H = \sum_{\alpha=1}^N H_\alpha\).
For example a Hamiltonian could be split into terms expressing hopping in \(x\)-direction, hopping in \(y\)-direction and on-site interaction.

The unitary transformation of the VHA is given by,
\begin{align}\label{eq:vha-evolution}
  U(\vec \theta) = \prod_{k=1}^n \prod_{\alpha=1}^N \mathrm{e}^{\mathrm i \theta_{\alpha,k} H_\alpha} \ ,
\end{align}
where the \(N\) terms of the Hamiltonian are applied \(n\)-times to the initial state in a pseudo-time-evolution.

The quality of the result of the variational approach depends on which states can be reached by the trial states. The set of possible trial states also determines which broken symmetry ground state can be found.

In general, the standard VHA cannot break symmetries of the Hamiltonian \(H\) that are also symmetries of the initial state \(\ket{\psi_0}\).
To illustrate this, consider a Hamiltonian $H$ that is invariant under a symmetry transformation \(\hat{T}_{\textrm{S}}\).
By definition the symmetry operator commutes with the full Hamiltonian,
\begin{equation}
  \left[\hat{T}_{\textrm{S}}, H \right] = 0 \ .
\end{equation}
In many cases of practical interest it will also commute with the separate terms of the Hamiltonian,
\begin{equation}
    \left[\hat{T}_{\textrm{S}}, H_{\alpha} \right] = 0 \ .
  \end{equation}
One example for the Hubbard model would be the spin-flip operator that is associated with the broken symmetry in the anti-ferromagnetic state.
The spin-flip operator commutes with all hopping terms and the interaction term of the Hubbard model separately.

The symmetry operator also commutes with all the unitary evolution operators that are applied to the initial state in the VHA algorithm individually in \eqref{eq:vha-evolution},
\begin{align}
  \hat{T}_{\textrm{S}} \left| \psi (\vec\theta) \right\rangle &= \hat{T}_{\textrm{S}} \prod_{s=1}^n \prod_{k=1}^m e^{-i \hat{H}_{\alpha} \theta_{\alpha,k}}\left|\psi_{0}\right\rangle \\
  &=  \prod_{s=1}^n \prod_{k=1}^m e^{-i \hat{H}_{\alpha} \theta_{\alpha,k}} \hat{T}_{\textrm{S}} \left|\psi_{0} \right\rangle \ .
\end{align}

If the initial state is invariant under the symmetry transformation, so is every state created by the VHA,
\begin{align}
  \hat{T}_{\textrm{S}} \left|\psi_{0}\right\rangle =  \left|\psi_{0}\right\rangle \quad \Rightarrow \quad \hat{T}_{\textrm{S}} \left| \psi_{\mathrm{vha}} \right\rangle = \left| \psi_{\mathrm{vha}} \right\rangle \ .
\end{align}

The only way to break a general symmetry \(\hat{T}_{\textrm{S}}\) of the Hamiltonian \(H\) in the standard VHA is to deliberately decompose the Hamiltonian into parts \(H_{\alpha}\) that do not individually commute with \(\hat{T}_{\textrm{S}}\).
We do not expect this to be a very efficient approach in general as one of the main advantages of the VHA is that the application of physically motivated unitary transformations to the initial states minimizes the number of classical parameters in the optimization problem.
Reverting to a very general decomposition of the Hamiltonian (for example each hopping term between two sites as a separate Hamiltonian) would bring back the scaling problems with system size that the VHA avoids compared to the general VQE.
Choosing a decomposition that only contains a small number of partial Hamiltonians but breaks symmetries on purpose would be at least as complex as using the VEHA (Sec:\ref{sec:VEHA}) and VMFHA (Sec:\ref{sec:VMFHA}) algorithms, but would lack the systematic approach to symmetry breaking of the VEHA and VMFHA.

\subsection{VHA with post-selection}
When using the VHA to obtain a ground state energy estimate for a system with known (broken) symmetries, the problem simplifies to preparing an initial state \(\ket{\psi_{0}}\) with the correct symmetry.
Additionally, the initial state should have a large overlap with the true ground state, in order to obtain the best possible estimate of the true ground state from the classical optimizer.
A good way to achieve a given broken symmetry and a large overlap with the true ground state is to use the ground state of a Mean Field approximation of the full interacting Hamiltonian as the initial state.
All ground states of fermionic quadratic Mean Field Hamiltonians are Gaussian states which can be efficiently prepared on a quantum computer with \(\mathcal{O}(M^2)\) gates and \(\mathcal{O}(M)\) circuit depth\cite{jiang_quantum_2018}, where \(M\) is the number of fermionic modes in the Hamiltonian.

Consider the general interacting, non-particle-number-conserving Hamiltonian,
\begin{align}
  H &= \sum_{ij} \alpha_{ij} c^{\dagger}_i c_j + \textrm{h.c.} + \sum_{ij} \beta_{ij} c_i c_j + \textrm{h.c.}\nonumber \\
  &+ \sum_{ijkl} \gamma_{ijkl} c^{\dagger}_i c^{\dagger}_j c_k c_l + \textrm{h.c.} \ ,
\end{align}
where \(c_i\) are fermionic annihilation operators. With the particle conserving mean field \(\Delta^\mathrm{c}_{ij}=\expval{\hat \Delta^\mathrm{c}_{ij}} = \expval{-\sum_{k,l} \gamma_{ikjl} c_{k}^{\dagger}  c_{l}}\) and the non-conserving field \(\Delta^\mathrm{nc}_{ij}=\expval{\hat \Delta^\mathrm{nc}_{ij}} = \expval{\sum_{k,l} \gamma_{ijkl} c_{k}^{\dagger}  c^{\dagger}_{l}}\), the corresponding mean field Hamiltonian reads,
\begin{align}
  \label{eq:MF_Ham}
  H_{MF} &= \sum_{ij} \alpha_{ij} c^{\dagger}_i c_j + \textrm{h.c.} + \sum_{ij} \beta_{ij} c_i c_j + \textrm{h.c.}  \nonumber \\
  &+ \sum_{ij} \Delta^{c}_{ij} c^{\dagger}_i c_j + \textrm{h.c.} + \sum_{ij} \Delta^{nc}_{ij} c_i c_j + \textrm{h.c.} \ .
\end{align}
The values of the mean fields are determined by a simple self-consistency loop, repeatedly evaluating the expectation values of \(\hat \Delta^{c/nc}_{ij}\) with respect to the the ground state of \(H_{MF}\) until the expectation values and the values in \(H_{MF}\) converge.

The (broken) symmetries of the Mean Field ground state depend on the Mean Field operators included at the start of the self-consistency loop.
For example, in a superconductor the chosen Mean Field would be the superconducting gap,
\begin{align}
  \Delta^{c}_{ij} &= 0 \\
  \Delta^{nc}_{ij} &= \delta_{j,i} U \left\langle \sum_{k,l} \delta_{k, l} \hat c_{k}^{\dagger} c_{l}^{\dagger}\right\rangle \ ,
\end{align}
where \(\delta_{j,i}\) and therefore \(\Delta^{nc}_{ij}\) are only non-zero when \(i\) and \(j\) correspond to states with opposite momentum and spin: \(i= (m, \uparrow)\) and \(i= (-m, \downarrow)\), where \(m\) is a momentum.

When the system is close to a phase transition and the correct order of the ground state is not known, we can still use Mean Field ground states as initial states.
We assume that we understand the system in question sufficiently well to know the possible order parameters and the corresponding Mean Field operators close to the phase transition.
We perform an independent VHA run for each possible order with the self consistent Mean Field ground state as the respective initial state.
From the set of results \(\{E_1, E_2, \ldots \}\) we choose the lowest energy in a post-selection.
In accordance with the variational principle \eqref{eq:var_p}, the energy and order chosen by post-selection are our best approximations of the true ground state energy and ground state wave function.

The post-selection approach is similar to the approach of Ref.\onlinecite{wecker_solving_2015}: start in a ground state of a symmetry breaking Hamiltonian, adiabatically evolve to the desired Hamiltonian and detect symmetry transitions on the way. When no symmetry transition occurs, the starting state had the correct symmetry of the ground state.

\subsection{Variational Extended Hamiltonian Ansatz}
\label{sec:VEHA}
As an alternative to post-selection we propose the Variational Extended Hamiltonian Ansatz (VEHA), where compared to the standard VHA in \eqref{eq:vha-evolution} additional explicitly symmetry breaking terms and optimization parameters \(\vec{\theta}^{\textrm{E}}\) are introduced. 
An approach using set symmetry breaking terms without additional variational parameters has recently been used in Ref.\onlinecite{endo_variational_2020}.
In the VEHA we start from an empty or vacuum state \(\ket{\textrm{vac}}\) and create particles in the system by applying the symmetry breaking terms.
We obtain the VEHA ansatz form the standard VHA \eqref{eq:vha-evolution} with the substitutions:
\begin{align}
  \ket{\psi_0} &\rightarrow \ket{\textrm{vac}} \\
  \{H_{\alpha}\} &\rightarrow \{H_{\alpha}, H^{\textrm{E}}_{\beta}\} \\
  \vec \theta &\rightarrow \vec \theta , \vec{\theta}^{\textrm{E}} \\
  U(\vec \theta ,  \vec{\theta}^{\textrm{E}}) &= \prod_{k=1}^n \left(\prod_{\alpha} \mathrm{e}^{\mathrm i \theta_{\alpha,k} H_\alpha} \right) \left( \prod_{\beta} \mathrm{e}^{\mathrm i \theta^{\textrm{E}}_{\beta,k} H_\beta^{\textrm{E}}} \right)\ .
\end{align}
The symmetry breaking terms are chosen based on the potential ground state phases of the system so that they would prepare a symmetry broken state starting from an empty system. These additional terms must always be able to at least break particle-number conservation to leave the vacuum state,
\begin{equation}
  \ket{\psi_{\textrm{sym-broken}}} = \mathrm{e}^{\mathrm i  H_1^{\textrm{E}}} \ket{\textrm{vac}} \ .
\end{equation}

For example, if we suspect that a given system has a superconducting ground state, we add the explicitly symmetry breaking term,
\begin{align}
  H_1^{\textrm{E}}=\sum_i c_{i \uparrow}^{\dagger} c_{i \downarrow}^{\dagger} + \textrm{h.c.} \ ,
\end{align}
that creates a Cooper pair on each site.
It is important to note that the explicitly symmetry breaking terms \(H^{\textrm{E}}_{\beta}\) are in general not the same as the order parameter operators \(\Delta^{c/nc}_{ij}\).

During the minimization of the energy, the classical optimizer determines how particles are added to the system via the parameters \(\vec{\theta}^{\textrm{E}}\). If the optimization succeeds, it finds a state with the correct symmetry.

In principle, non-particle number conserving terms could also be added to a VHA with a traditional initialization to try to change the symmetry of the initial state.
In this case the optimizer needs to find a set of parameters that first removes population from the system and then adds population in a way that changes the symmetry.
This is difficult and the VEHA encounters instabilities in the optimizer when used on top of some initial state already containing fermions or bosons.
The VEHA is best used starting from the empty state.

\subsection{Variational Mean Field Hamiltonian Ansatz}
\label{sec:VMFHA}
The Variational Mean Field Hamiltonian Ansatz (VMFHA) is a combination of the VHA with post-selection and the VEHA.
It starts from a Mean Field ground state, but allows the classical optimizer to influence the initial mean fields.

We make an ansatz for a Mean Field Hamiltonian as in \eqref{eq:MF_Ham}, including all Mean Fields \(\Delta^{c/nc}_{ij}\) that correspond to
potential orders of the system.
For the transition between superconducting and anti-ferromagnetic order in the Hubbard model, for example, we would choose the superconducting gap and the anti-ferromagnetic order.
The values of the Mean Fields are not determined by a self-consistency loop but taken to be free parameters in the initialization,
\begin{align}
  \ket{\psi( \vec\theta , \{\Delta^{c}_{ij}\}, \{\Delta^{nc}_{ij}\})} &= U(\vec\theta) \ket{\psi_0( \{\Delta^{c}_{ij}\}, \{\Delta^{nc}_{ij}\})} \ .
\end{align}
In the energy minimization, the Mean Fields are treated as classical optimization parameters,
\begin{align}
  E(\vec\theta_\mathrm{min},  \{\Delta^{c/nc}_{ij}\}_\mathrm{min}) = \min_{\vec\theta, \{\Delta^{c/nc}_{ij}\}} E(\vec\theta , \{\Delta^{c/nc}_{ij}\}) \ .
\end{align}

Including the Mean Fields in the optimization allows the optimizer to find the dominant Mean Field and even solutions with non-zero expectation values for several mean-fields.

The optimization of the initial state preparation in the VMFHA is similar to the approach recently used by Google to run the Hartree-Fock method on a quantum computer \cite{arute_hartree-fock_2020}

\subsection{Circuit depths of the Variational Algorithms}
\label{sec:circ_depth}
The circuit depth of the VHA and VMFHA is determined by two parts of the algorithm, the initial state preparation and the unitary evolution of the initial state in the trial state preparation.
Following Ref.\onlinecite{jiang_quantum_2018} the circuit depth of the initialization scales as \(\mathcal{O}(N)\) with system size \(N\) while the total number of gates scales as \(\mathcal{O}(N^2)\).
The scaling of the unitary evolution part strongly depends on the available quantum computer topology and the type of system Hamiltonian that is decomposed into partial Hamiltonians \(H_{\alpha}\).
We cannot make a general statement about the scaling behaviour, but taking the two-dimensional Hubbard model as an example, the unitary evolution of the system under the full Hubbard Hamiltonian can be calculated with \(\mathcal{O}(\sqrt{N})\) circuit depth and \(\mathcal{O}(N)\) gates.
Note that not every unitary evolution under a partial Hamiltonian \(H_{\alpha}\) requires that circuit depth as the decomposed Hamiltonians can be easier to simulate than the full Hubbard Hamiltonian.

In the VEHA, the circuit depth is fully determined by the unitary evolution as we start from an empty system without initialization.
The unitary evolution will have slightly higher depth as it now contains the additional evolution corresponding to the \(H^{\textrm{E}}_{\beta}\) terms.
Again we cannot make general scaling statements, but for the example of a two-dimensional Hubbard model the additional depth due to the \(H^{\textrm{E}}_{\beta}\) terms is in \(\mathcal{O}(1)\) while the number of additional gates scales as  \(\mathcal{O}(N)\).

\section{The Hubbard model as a test case}
\label{sec:hubbard}
To compare VHA with postselection, VEHA and VMFHA, we consider symmetry breaking in the two-dimensional Hubbard model.
The system sizes accessible to numerical simulation of the full quantum algorithm are too small to observe proper symmetry breaking.
Therefore, instead of a full phase transition, we consider the change in the dominant response of two characteristic expectation values to external symmetry breaking fields.
Analyzing a ground state with an explicitly broken symmetry is sufficient to study broken symmetry state preparation with the Variational Hamiltonian Ansatz.
The two expectation values we consider are the s-wave BCS gap \(\Delta_s\) and the anti-ferromagnetic magnetization \(M_\mathrm{AF}\),
\begin{align}
  \label{eqn:Hubbard_order_parameters}
  M_\mathrm{AF} &= \left\langle \frac{1}{N_x N_y} \sum_{x,y, \sigma} \hat{\alpha}_\mathrm{AF}(x,y,\sigma) \right\rangle \\
  \Delta_s &= \left\langle \sum_{x,y} U \hat{v}_{s}(x,y) \right\rangle \ ,
\end{align}
where the anti-ferromagnetic (\(\hat{\alpha}_\mathrm{AF}(x,y,\sigma)\)) coupling operator is defined as,
\begin{align}
    \label{eqn:AF_coupling_op}
    \hat{\alpha}_\mathrm{AF}(x,y,\sigma) &= \left(-1\right)^{p(x,y)} \times \sigma \times \hat{n}_{x,y,\sigma} \\
    p(x,y) &= \mod(x+y, 2) \ ,
  \end{align}
and the  superconducting \(\hat{v}_{s}(x,y)\) operator as,
\begin{align}
    \label{eqn:supercond_coupling_op}
    \hat{v}_{s}(x,y) &= c_{x,y, \downarrow} c_{x,y, \uparrow} \ .
\end{align}

We simultaneously apply an external anti-ferromagnetic field \(B_\mathrm{AF}^{\textrm{ext}}\) and an external superconducting gap \(\Delta_s^{\textrm{ext}}\) explicitly breaking both spin flip/translational and \(U(1)\) symmetry.
With coupling to external fields the Hubbard model Hamiltonian reads,
\begin{align}
  H = & \sum_{\left< (x,y),(x',y') \right>,\sigma} t \ c_{x,y,\sigma}^{\dagger} c_{x',y',\sigma} + \textrm{h.c.} \\
  +& \sum_{x,y} U \left(\hat{n}_{x,y, \uparrow }- \frac{1}{2}\right) \left(\hat{n}_{x,y, \downarrow} - \frac{1}{2}\right) \\
  +& \sum_{x,y, \sigma} B_\mathrm{AF}^{\textrm{ext}} \hat{\alpha}_\mathrm{AF}(x,y,\sigma) \\
  +& \sum_{x,y} \Delta_s^{\textrm{ext}} \hat{v}_{s}(x,y) + \textrm{h.c.} \ ,
  \label{eqn:Hub_model_expl_3d}
\end{align}
where we assume a discrete lattice with \(x,y \in \left[0, N_{x,y} -1 \right] \).

For all Variational Hamiltonian Ansätze we decompose the Hubbard Hamiltonian without external fields into hopping in x-direction for even and odd sites, hopping in y-direction for even and odd sites and interaction,
\begin{align}
  H &= \sum_{\alpha=1}^5 H_\alpha \\
  H_1 &=\sum_{x,y \textrm{ for } x \textrm{ odd},\sigma} t \ c_{x,y,\sigma}^{\dagger} c_{x+1,y,\sigma} + \textrm{h.c.}  \\
  H_2 &=\sum_{x,y \textrm{ for } x \textrm{ even},\sigma} t \ c_{x,y,\sigma}^{\dagger} c_{x+1,y,\sigma} + \textrm{h.c.}  \\
  H_3 &=\sum_{x,y \textrm{ for } y \textrm{ odd},\sigma} t \ c_{x,y,\sigma}^{\dagger} c_{x,y+1,\sigma} + \textrm{h.c.}  \\
  H_4 &=\sum_{x,y \textrm{ for } y \textrm{ odd},\sigma} t \ c_{x,y,\sigma}^{\dagger} c_{x,y+1,\sigma} + \textrm{h.c.}  \\
  H_5 &= \sum_{x,y} U \left(\hat{n}_{x,y, \uparrow }- \frac{1}{2}\right) \left(\hat{n}_{x,y, \downarrow} - \frac{1}{2}\right) \ .
\end{align}

For VHA with post-selection and VMFHA we consider two types of Mean Field Hamiltonians: the superconducting model where the mean-field parameter is simply the BCS gap \(\Delta_s\) and the anti-ferromagnet with two Mean Fields corresponding to the occupation of up and down spin orbitals on the even and odd sites of the lattice,
\begin{align}
  \label{eqn:AF_order_parameters}
  n_{p=-1} & = \left\langle \sum_{\substack{(x,y,\sigma) \\ p(x,y,\sigma)=-1}} \hat{n}_{x,y, \sigma}\right\rangle \\
  n_{p=1} & = \left\langle \sum_{\substack{(x,y) \\ p(x,y,\sigma)=1}} \hat{n}_{x,y, \sigma}\right\rangle \\
  p(x,y,\sigma) &= \sigma \times (-1)^{\textrm{mod}(x +y, 2)} \ .
\end{align}

In the VEHA we use three combinations of symmetry breaking terms \(\{H_\mathrm{BCS}\}\), \(\{H_{AF, p=\pm1 }\}\) and \(\{H_\mathrm{BCS}, H_{AF, p=\pm1 }\}\), where
\begin{align}
    \label{eqn:symm_breaking_terms}
    H_\mathrm{BCS} &= \sum_{x,y,z} \left(c_{x,y,z, \downarrow} c_{x,y,z, \uparrow} + c^{\dagger}_{x,y,z, \uparrow} c^{\dagger}_{x,y,z, \downarrow} \right)\\
    H_{AF, p=-1} &= \sum_{\substack{(x,y,\sigma) \\ p(x,y,\sigma)=-1}} \left(c_{x,y, \sigma}  + c^{\dagger}_{x,y, \sigma} \right) \\
    H_{AF, p=1} &= \sum_{\substack{(x,y,\sigma) \\ p(x,y,\sigma)=1}} \left(c_{x,y, \sigma}  + c^{\dagger}_{x,y, \sigma} \right) \ .
  \end{align}
Additionally, we apply post-selection based on the energy found in each case in the same way we do for the VHA with post-selection.

\section{Numerical results}
\label{sec:numeric}

In this section we show the numerical results of simulated runs of the VHA with post-selection, the VEHA and the VMFHA.
For all simulations the number of electrons in the Hubbard model is set to half filling via the initialization for the VHA and the VMFHA or via the chemical potential for the VEHA. The hopping is set to \(t=-1\) for all simulations.

We scale the external BCS and anti-ferromagnetic fields with \(U\), but introduce a minimum field strength to avoid vanishing external fields close to \(U=0\)
\begin{align}
  \Delta_s^{\textrm{ext}} &= \max(0.1, 0.1 U) \\
  B_\mathrm{AF}^{\textrm{ext}} =  &= \max(0.1, 0.1 U) \ .
\end{align}

The initialization of \(\ket{\psi_0}\) and the unitary transformations \(U_i(\theta_i)\) are simulated with a full gate-based simulator of a quantum computer. We use the QuEST\cite{jones_quest_2019} simulator via the PyQuEST-cffi python interface.
We assume a quantum computer with \gate{RX}, \gate{RY} and \gate{RZ} single qubit gates and a controlled Pauli Z (\gate{CZ}) two qubit entangling gate.
For simplicity we assume a fully connected qubit topology and use a simple \gate{CZ}-algorithm \cite{reiner_finding_2018} for the trotterization of the time-evolution of the Hubbard model.
While such a topology is hard to realize in solid state systems like superconducting qubits\cite{nakamura_coherent_1999, clarke_superconducting_2008}, quantum dots\cite{zwanenburg2013silicon, veldhorst2014addressable, watson2018programmable} or doped silicon\cite{he2019two}, it is common in trapped ion devices\cite{cirac_quantum_1995, bermudez_assessing_2017}.
Even for architectures without a fully connected qubit topology similarly short quantum circuits can be realized by using advanced trotterization algorithms\cite{kivlichan_quantum_2018}.

In our simulations the measurement is performed by matrix multiplication with the simulated state vector for numerical efficiency.
This allows us to study the ideal performance of the algorithms without statistical fluctuations due to a finite number of measurements.
In the appendix we also show a simulation with measurements by one-qubit rotations and projective measurements of the qubits in the \(\sigma^z\) basis.

The initialization of Mean Field ground states is performed with Givens rotations and particle hole transformations\cite{jiang_quantum_2018}, where the correct sequence is obtained from the openfermion package\cite{mcclean_openfermion_2019}.
We trotterize the unitary evolutions \(e^{\mathrm i H_\alpha \theta}\) with respect to \(H_{\alpha}\) to second order and use \(n=4\) repetitions of the list of partial Hamiltonians \(\{H_\alpha\}\) \eqref{eq:vha-evolution}.

\subsection{The ideal case}
\begin{figure}
  \centering
  \includegraphics[width=.5\textwidth]{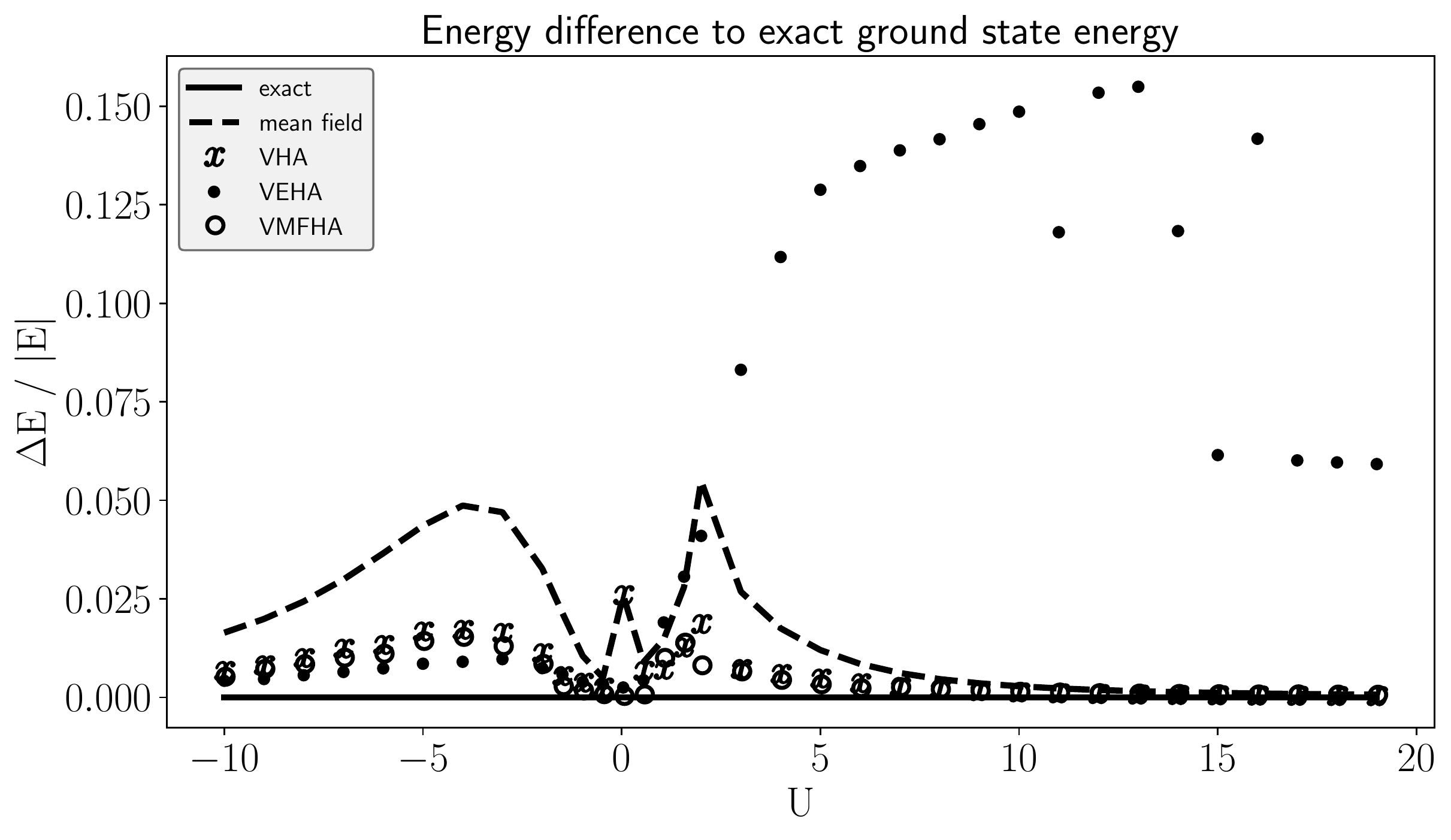}
  \caption{%
      \label{fig:3_2_E}
      The energy difference between the calculated and the exact ground state energy scaled to the absolute value of the exact ground state energy for the Mean Field solver (black dashed line), the VHA with post-selection (black x), the VEHA (black circles) and the VMFHA (black cross) as a function of the interaction strength. The variational algorithms improve upon the Mean Field result except for the VEHA algorithm that shows an instability for positive \(U\). The improvement compared to standard Mean Field results is to be expected for the VHA and VMFHA as they start form the Mean Field results and the classical optimizer in the algorithm specifically minimizes energy. Overall VMFHA shows the best results.
  }%
\end{figure}
\begin{figure}
  \centering
  \includegraphics[width=.5\textwidth]{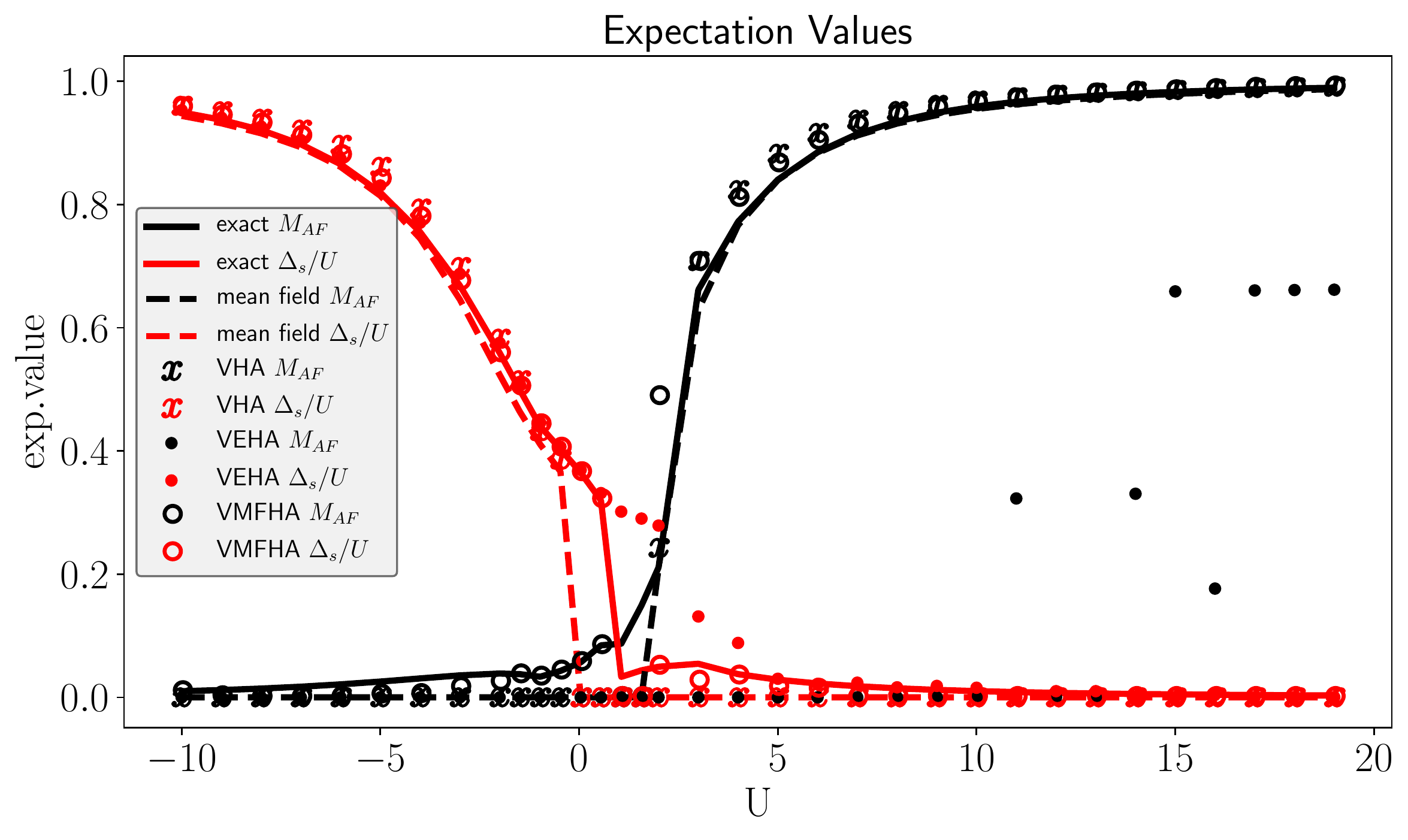}
  \caption{%
      \label{fig:3_2_exp_val}
      The superconducting gap \(\Delta_s\) (red) and the antiferromagnetic magnetization \(M_\mathrm{AF}\) (black) for exact diagonalization (solid line), the Mean Field solver (dashed line), the VHA with post-selection (x), the VEHA (dots) and the VMFHA (circles) as a function of the interaction strength. Classical Mean Field, VHA and VMFHA show the transition in the dominant expectation value.
       The VEHA exhibits good results in the region dominated by \(\Delta_s\), but fails to reproduce the dominant \(M_\mathrm{AF}\) correctly at positive \(U\) values.
      At small positive \(U\) the VHA results for \(M_\mathrm{AF}\) (\(x\)) are further away from the exact solution (solid line) than the pure Mean Field results (dashed line).
      Overall the VMFHA produces the best results, partially reproducing the linear response in the non-dominant expectation value that is not reproduced by the classical Mean Field solver, the VHA or the VEHA.
  }%
\end{figure}
In the ideal case simulation we neglect any effects from physical noise and the finite number of measurements.
We consider a \(3 \times 2\) Hubbard model with periodic boundary conditions and compare the results of the variational algorithms to the exact expectation values obtained by matrix diagonalization and the Mean Field expectation values obtained form the self-consistent loop.
We use scipy's gradient based ``L-BFGS-B'' optimizer\cite{lbfgsb} for the classical minimization of the energy and repeat each run \(10\) times with different randomly chosen initial values for the optimizer, post-selecting for the lowest energy result.
We vary the interaction strength \(U\) from negative to positive values.
In Fig.\ref{fig:3_2_E} we show the energy differences between the ground state energy found by the different algorithms and the ground state energy found by exact diagonalization.
All variational algorithms improve upon the standard Mean Field result except for the VEHA at positive \(U\). The optimizer in the variational algorithms minimizes the energy. The improvement in the ground state energy approximation compared to standard Mean Field theory is the expected behaviour for all variational algorithms that start from the Mean Field result.

While there is no phase transition between the BCS and the anti-ferromagnetic phase at \(U=0\) in Fig.\ref{fig:3_2_exp_val}, we see a change in the dominant expectation value from the superconducting gap \(\Delta_s\) at negative \(U\) to the anti-ferromagnetic magnetization \(M_\mathrm{AF}\) for large positive \(U\).
The standard Mean Field results, the VHA with postselection and the VMFHA show good agreement with the exact solution for the dominant expectation value except for a small region around the transition between dominant expectation values.

The VEHA shows good agreement as long as \(\Delta_s\) is dominant, but is unstable for larger positive \(U\).
The additional operators in the VEHA ansatz generally break the \(U(1)\) symmetry, which is not a problem in the BCS-dominated regime \( U < 0 \) as we are searching for a ground state with broken \(U(1)\) symmetry.
In the AF-regime, however, the symmetry is not broken and restoring the \(U(1)\) symmetry requires an exact balancing of the optimized parameters in the VEHA.
This makes the optimization problem for repulsive \(U\) much harder to solve numerically.
The numerical simulation results do not provide enough data to decide whether this problem is too hard for the chosen optimizer or  so fundamental, that the VEHA cannot converge for repulsive \(U\) in general.
While this is an interesting topic for further research, a detailed comparison of classical optimizers in the context of the VEHA would go beyond the scope of this paper.

Overall the VMFHA shows the best performance out of the three compared algorithms.
It can also partially reproduce the linear response of the non-dominant expectation value to the external fields for small absolute values of \(U\).

\subsection{Decoherence}
\begin{figure}
  \centering
  \includegraphics[width=.5\textwidth]{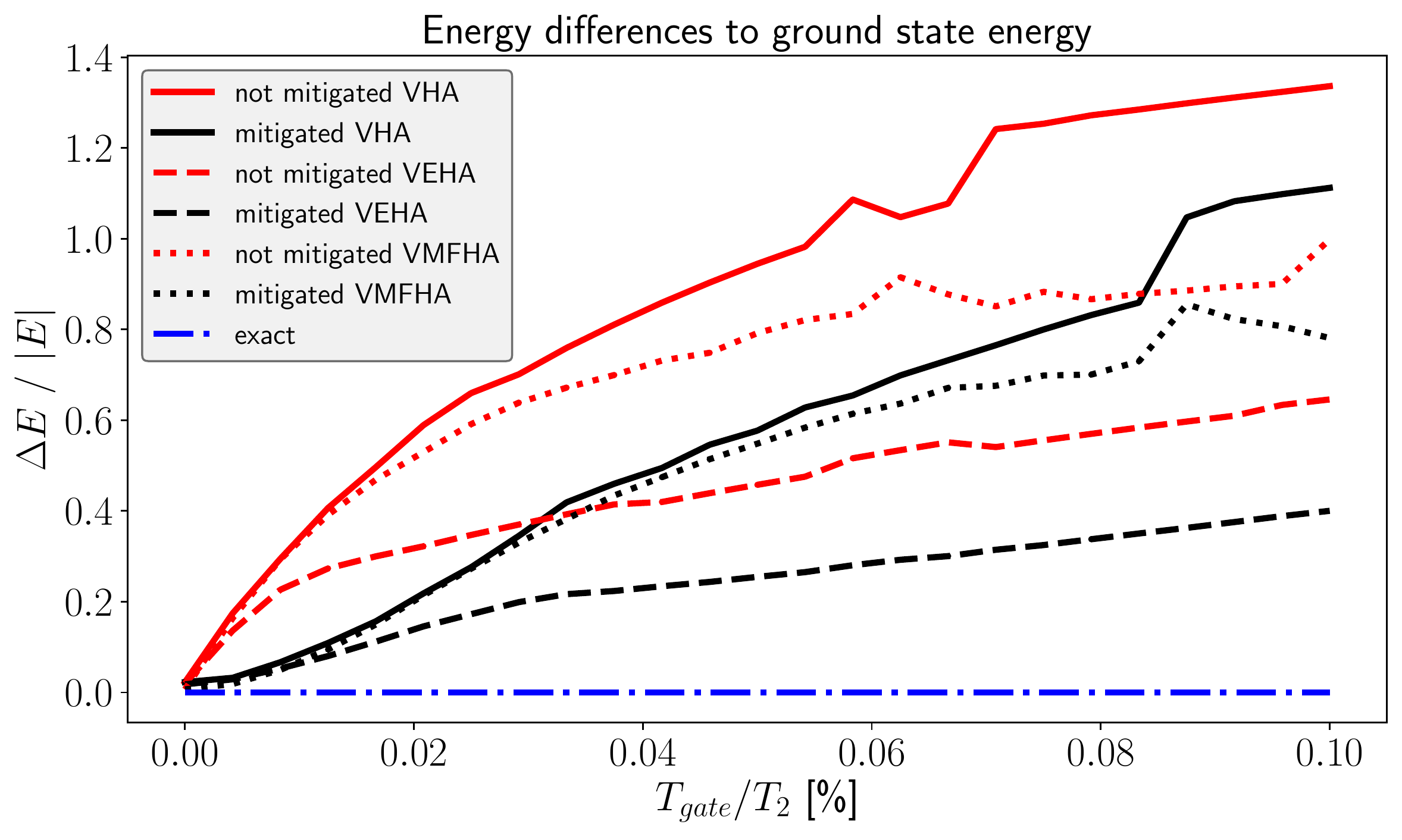}
  \caption{\label{fig:2_2_E}
      The energy difference between the calculated and the exact ground state energy scaled to the absolute value of the exact ground state energy for unmittigated (red) and mitigated (black) dephasing, for the VHA with post-selection (solid line), the VEHA (dashed line) and the VMFHA (dotted line) as a function of gate time over dephasing time at \(U = -3 \). With increasing dephasing the deviation of the quantum algorithms from the exact result grows as expected, reaching up to \(1.4\) times the exact energy value at the longest gate time of \(10^{-4} T_2\). Richardson mitigation improves the results significantly for all algorithms. In this regime, dominated by the BCS response, the VEHA is stable and performs better than the other algorithms with increasing noise.
  }
\end{figure}
\begin{figure}
  \centering
  \includegraphics[width=.5\textwidth]{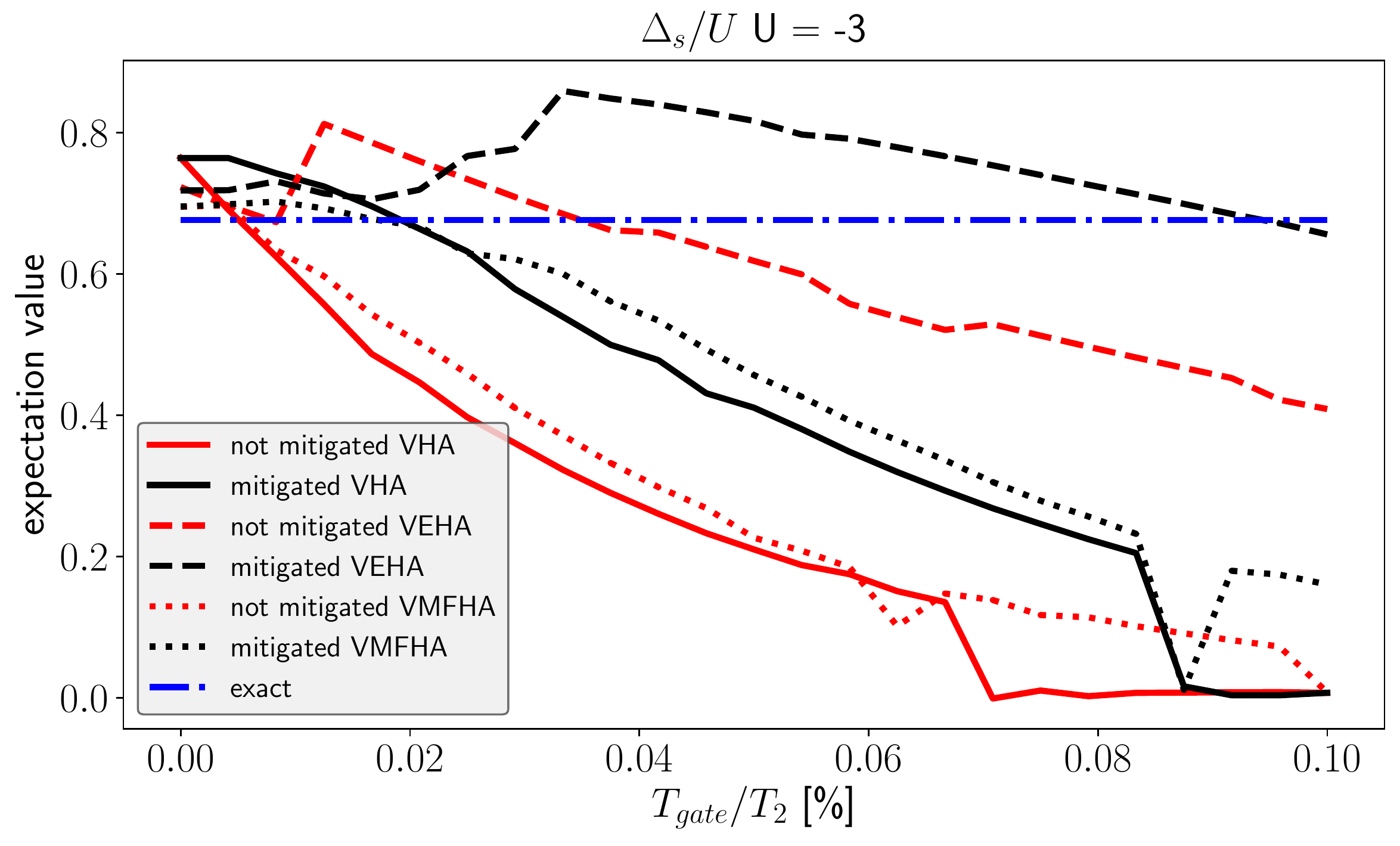}
  \caption{%
      \label{fig:2_2_exp_val_D}
      The superconducting gap \(\Delta_s\) for exact diagonalization (blue dotted-dashed line), for unmittigated (red) and mitigated (black) dephasing, for the VHA with post-selection (solid line), the VEHA (dashed line) and the VMFHA (dotted line) as a function of gate time over dephasing time at \(U=-3\). The results for \(\Delta_s\) confirm the observations from Fig.\ref{fig:2_2_E}, and deteriorate with increasing noise. In the BCS regime, where the VEHA is stable, it produces the best results and mitigation improves results for all algorithms.
  }%
\end{figure}

To study the effects of decoherence on the variational algorithms we consider a \(2 \times 2\) Hubbard model with periodic boundary conditions in the presence of pure dephasing noise on each qubit.
Compared to the decoherence free case, simulations including decoherence require significantly more computational resources and we reduce the system size to \(2 \times 2\) to compensate this effect.
For simplicity we limit the simulations to pure dephasing, but we do not expect qualitatively different results when including qubit damping or depolarisation.

Gradient based solvers are not suited for the noisy simulations.
When numerically determining the gradients, small fluctuations in the energy measurement due to decoherence lead to unrealistically large gradients and the gradient based solver does not converge.
An alternative are black-box solvers, which do not calculate numerical gradients and are not affected by this problem.
We choose the widely used ``COBYLA'' black-box optimizer from the scipy python package for the classical minimization of the energy. Since minimization results depend on the initial values of the optimized parameters, we repeat each run \(10\) times with different randomly chosen initial values for the optimizer, post-selecting for the lowest energy result.

We assume uniform noise that is determined by the qubit dephasing time (commonly referred to as \(T_2\) time in literature) and given in multiples of the gate time \(T_{gate}\) of the single qubit gates \gate{RX} and \gate{RY}.
The \gate{RZ} gate can be implemented by shifting the energy gap between \(\ket{0}\) and \(\ket{1}\) qubit states.
We assume that this operation is fast enough on the quantum computer (as we would expect in superconducting qubits) that the decoherence during the \gate{RZ} gate can be neglected.
For two-qubit entangling gates, we assume the \gate{CZ} gate takes three times as long as \gate{RX} and \gate{RY}.
While we chose this set of parameters, our choice does not represent any specific available quantum computing device.
For this general study of the effects of noise we simply need a way to define the single and two-qubit fidelities as functions of one parameter (\(T_2\)).

For all algorithms we compare simulations with noise with simulations with mitigated noise.
For the noise mitigation we use Richardson extrapolation \cite{Temme_extrapolation_17}, where the noise-free results are extrapolated from a fit to measurements with the base noise level and increased noise levels.
We assume that the noise can be increased by \(50\%\) and use two datapoints for a linear extrapolation.
One way that noise could be increased in a physical device is to increase the gate time of all gates in the circuit by \(50\%\).
How well gate times can be controlled will strongly depend on the given architecture of a quantum computing device.
The results of mitigation might be further improved when using more datapoints for the extrapolation.

The effects of stochastic fluctuations in the measurement results due to a finite number of measurements can be reduced by increasing the number of measurements. The stochastic fluctuations decrease with the square root of the number of measurements.
Here we consider the limit of infinite projective measurements and use perfect measurements based on matrix multiplication in the calculations.
The effects of a finite number of measurements (\(25000\)) are shown in appendix \ref{sec:stochastic}.

We analyze different noise strengths by varying the gate time in relation to the \(T_2\) time from \(0\), corresponding to the noise free case, to \(0.1 \% T_2\).
The upper bound of the dephasing strength was chosen such that the decay of the expectation values from the exact solution with increasing dephasing strength occurs on the range of dephasing strengths shown in the figures.
The largest dephasing rate we consider is still smaller than what has been recently demonstrated by the breakthrough experiment \onlinecite{arute_quantum_2019} but the difference is less than an order of magnitude.

In Fig.\ref{fig:2_2_E} and Fig.\ref{fig:2_2_exp_val_D} we show how the results of all variational algorithms deteriorate with increasing dephasing.
Since we explicitly want to compare the performance of the three considered algorithms with increasing dephasing strength, we choose the parameter regime \(U=-3\) where the VEHA is stable.
For the anti-ferromagnetic parameter regime we expect no qualitative difference for the VHA and VMFHA and that the VEHA is again unstable.
We note that Richardson mitigation significantly improves the results.
The VEHA shows the best performance of all algorithms in the parameter regime dominated by \(\Delta_s\) where it is stable.
It is less affected by noise because the required quantum circuits are shorter. This is because it starts from an empty system and does not require the long initialization circuit preparing the ground state of a Mean Field Hamiltonian as discussed in Sec\ref{sec:circ_depth}.
\section{Conclusion}
In this work we study the performance of three variants of the Variational Hamiltonian Ansatz (VHA) designed to find ground states with a broken symmetry, the Variational Hamiltonian Ansatz with post-selection, the Variational Extended Hamiltonian Ansatz (VEHA) and the Variational Mean Field Hamiltonian Ansatz (VMFHA).

We simulate all three algorithms running on a gate-based quantum computer to find the ground state of the two-dimensional Hubbard model.
As only small system sizes are numerically feasible, we explicitly break symmetries with external fields (an external superconducting gap and an external anti-ferromagnetic field) and observe a transition from a dominant response in the superconducting gap expectation value \(\Delta_s\) at negative interaction strength \(U\) to a dominant response in the anti-ferromagnetic magnetization \(M_\mathrm{AF}\) at large positive \(U\).

We show that the VHA with post-selection and the VMFHA are in good agreement with the exact results and reproduce the change in the dominant response in the expectation values.
The VEHA is in agreement with the exact solution when \(\Delta_s\) dominates, but is unstable in the \(M_\mathrm{AF}\) dominated regime.
Overall the VMFHA shows the best performance and stability in finding the approximate ground state energy and match the expectation values of the exact solution in the ideal case.

We also consider the effect of pure dephasing noise on the algorithms, varying the noise strength via the gate time of the quantum computer time from \(0\) to \(10^{-4} \, T_2 \).
As expected the quality of the results decreases with increasing noise.
We show that applying Richardson mitigation in the variational algorithms improves the results significantly.
We conclude that Richardson mitigation should be used when possible.
In the regime \(U < 0\), where the VEHA is stable, it performs best with increasing noise.
This result cannot be generalized for the regime \(U > 0 \), where the VEHA is unstable in the decoherence-free case.

We conclude that all three algorithms are suited to find broken symmetry ground states, with a preference for the VMFHA for low noise situations and \(U >0 \) and a preference for the VEHA in noisy systems for \(U < 0\).

\acknowledgements
We thank Peter Schmitteckert for insightful discussions about numerical simulations and the Hubbard model.

\bibliography{variational_broken_symmetry.bib}

\begin{thebibliography}{28}%
\makeatletter
\providecommand \@ifxundefined [1]{%
 \@ifx{#1\undefined}
}%
\providecommand \@ifnum [1]{%
 \ifnum #1\expandafter \@firstoftwo
 \else \expandafter \@secondoftwo
 \fi
}%
\providecommand \@ifx [1]{%
 \ifx #1\expandafter \@firstoftwo
 \else \expandafter \@secondoftwo
 \fi
}%
\providecommand \natexlab [1]{#1}%
\providecommand \enquote  [1]{``#1''}%
\providecommand \bibnamefont  [1]{#1}%
\providecommand \bibfnamefont [1]{#1}%
\providecommand \citenamefont [1]{#1}%
\providecommand \href@noop [0]{\@secondoftwo}%
\providecommand \href [0]{\begingroup \@sanitize@url \@href}%
\providecommand \@href[1]{\@@startlink{#1}\@@href}%
\providecommand \@@href[1]{\endgroup#1\@@endlink}%
\providecommand \@sanitize@url [0]{\catcode `\\12\catcode `\$12\catcode
  `\&12\catcode `\#12\catcode `\^12\catcode `\_12\catcode `\%12\relax}%
\providecommand \@@startlink[1]{}%
\providecommand \@@endlink[0]{}%
\providecommand \url  [0]{\begingroup\@sanitize@url \@url }%
\providecommand \@url [1]{\endgroup\@href {#1}{\urlprefix }}%
\providecommand \urlprefix  [0]{URL }%
\providecommand \Eprint [0]{\href }%
\providecommand \doibase [0]{http://dx.doi.org/}%
\providecommand \selectlanguage [0]{\@gobble}%
\providecommand \bibinfo  [0]{\@secondoftwo}%
\providecommand \bibfield  [0]{\@secondoftwo}%
\providecommand \translation [1]{[#1]}%
\providecommand \BibitemOpen [0]{}%
\providecommand \bibitemStop [0]{}%
\providecommand \bibitemNoStop [0]{.\EOS\space}%
\providecommand \EOS [0]{\spacefactor3000\relax}%
\providecommand \BibitemShut  [1]{\csname bibitem#1\endcsname}%
\let\auto@bib@innerbib\@empty
\bibitem [{\citenamefont {Peruzzo}\ \emph {et~al.}(2014)\citenamefont
  {Peruzzo}, \citenamefont {McClean}, \citenamefont {Shadbolt}, \citenamefont
  {Yung}, \citenamefont {Zhou}, \citenamefont {Love}, \citenamefont
  {Aspuru-Guzik},\ and\ \citenamefont {O'Brien}}]{peruzzo_variational_2014}%
  \BibitemOpen
  \bibfield  {author} {\bibinfo {author} {\bibfnamefont {A.}~\bibnamefont
  {Peruzzo}}, \bibinfo {author} {\bibfnamefont {J.}~\bibnamefont {McClean}},
  \bibinfo {author} {\bibfnamefont {P.}~\bibnamefont {Shadbolt}}, \bibinfo
  {author} {\bibfnamefont {M.-H.}\ \bibnamefont {Yung}}, \bibinfo {author}
  {\bibfnamefont {X.-Q.}\ \bibnamefont {Zhou}}, \bibinfo {author}
  {\bibfnamefont {P.~J.}\ \bibnamefont {Love}}, \bibinfo {author}
  {\bibfnamefont {A.}~\bibnamefont {Aspuru-Guzik}}, \ and\ \bibinfo {author}
  {\bibfnamefont {J.~L.}\ \bibnamefont {O'Brien}},\ }\href {\doibase
  10.1038/ncomms5213} {\bibfield  {journal} {\bibinfo  {journal} {Nat Commun}\
  }\textbf {\bibinfo {volume} {5}},\ \bibinfo {pages} {4213} (\bibinfo {year}
  {2014})},\ \bibinfo {note} {arXiv: 1304.3061}\BibitemShut {NoStop}%
\bibitem [{\citenamefont {Wecker}\ \emph
  {et~al.}(2015{\natexlab{a}})\citenamefont {Wecker}, \citenamefont {Hastings},
  \citenamefont {Wiebe}, \citenamefont {Clark}, \citenamefont {Nayak},\ and\
  \citenamefont {Troyer}}]{wecker_solving_2015}%
  \BibitemOpen
  \bibfield  {author} {\bibinfo {author} {\bibfnamefont {D.}~\bibnamefont
  {Wecker}}, \bibinfo {author} {\bibfnamefont {M.~B.}\ \bibnamefont
  {Hastings}}, \bibinfo {author} {\bibfnamefont {N.}~\bibnamefont {Wiebe}},
  \bibinfo {author} {\bibfnamefont {B.~K.}\ \bibnamefont {Clark}}, \bibinfo
  {author} {\bibfnamefont {C.}~\bibnamefont {Nayak}}, \ and\ \bibinfo {author}
  {\bibfnamefont {M.}~\bibnamefont {Troyer}},\ }\href {\doibase
  10.1103/PhysRevA.92.062318} {\bibfield  {journal} {\bibinfo  {journal}
  {Physical Review A}\ }\textbf {\bibinfo {volume} {92}} (\bibinfo {year}
  {2015}{\natexlab{a}}),\ 10.1103/PhysRevA.92.062318},\ \bibinfo {note} {arXiv:
  1506.05135}\BibitemShut {NoStop}%
\bibitem [{\citenamefont {Nielsen}\ and\ \citenamefont
  {Chuang}(2011)}]{nielsen_quantum_2011}%
  \BibitemOpen
  \bibfield  {author} {\bibinfo {author} {\bibfnamefont {M.~A.}\ \bibnamefont
  {Nielsen}}\ and\ \bibinfo {author} {\bibfnamefont {I.~L.}\ \bibnamefont
  {Chuang}},\ }\href@noop {} {\emph {\bibinfo {title} {Quantum {Computation}
  and {Quantum} {Information}: 10th {Anniversary} {Edition}}}},\ \bibinfo
  {edition} {10th}\ ed.\ (\bibinfo  {publisher} {Cambridge University Press},\
  \bibinfo {address} {New York, NY, USA},\ \bibinfo {year} {2011})\BibitemShut
  {NoStop}%
\bibitem [{\citenamefont {Wecker}\ \emph
  {et~al.}(2015{\natexlab{b}})\citenamefont {Wecker}, \citenamefont
  {Hastings},\ and\ \citenamefont {Troyer}}]{Wecker_vha_15}%
  \BibitemOpen
  \bibfield  {author} {\bibinfo {author} {\bibfnamefont {D.}~\bibnamefont
  {Wecker}}, \bibinfo {author} {\bibfnamefont {M.~B.}\ \bibnamefont
  {Hastings}}, \ and\ \bibinfo {author} {\bibfnamefont {M.}~\bibnamefont
  {Troyer}},\ }\href {\doibase 10.1103/PhysRevA.92.042303} {\bibfield
  {journal} {\bibinfo  {journal} {Phys. Rev. A}\ }\textbf {\bibinfo {volume}
  {92}},\ \bibinfo {pages} {042303} (\bibinfo {year}
  {2015}{\natexlab{b}})}\BibitemShut {NoStop}%
\bibitem [{\citenamefont {Romero}\ \emph {et~al.}(2019)\citenamefont {Romero},
  \citenamefont {Babbush}, \citenamefont {McClean}, \citenamefont {Hempel},
  \citenamefont {Love},\ and\ \citenamefont {Aspuru-Guzik}}]{Romero_ucc_19}%
  \BibitemOpen
  \bibfield  {author} {\bibinfo {author} {\bibfnamefont {J.}~\bibnamefont
  {Romero}}, \bibinfo {author} {\bibfnamefont {R.}~\bibnamefont {Babbush}},
  \bibinfo {author} {\bibfnamefont {J.~R.}\ \bibnamefont {McClean}}, \bibinfo
  {author} {\bibfnamefont {C.}~\bibnamefont {Hempel}}, \bibinfo {author}
  {\bibfnamefont {P.~J.}\ \bibnamefont {Love}}, \ and\ \bibinfo {author}
  {\bibfnamefont {A.}~\bibnamefont {Aspuru-Guzik}},\ }\href {\doibase
  10.1088/2058-9565/aad3e4} {\bibfield  {journal} {\bibinfo  {journal} {Quantum
  Science and Technology}\ }\textbf {\bibinfo {volume} {4}},\ \bibinfo {pages}
  {1} (\bibinfo {year} {2019})},\ \Eprint
  {http://arxiv.org/abs/arXiv:1701.02691v2} {arXiv:arXiv:1701.02691v2}
  \BibitemShut {NoStop}%
\bibitem [{\citenamefont {McClean}\ \emph {et~al.}(2016)\citenamefont
  {McClean}, \citenamefont {Romero}, \citenamefont {Babbush},\ and\
  \citenamefont {Aspuru-Guzik}}]{mcclean_theory_2016}%
  \BibitemOpen
  \bibfield  {author} {\bibinfo {author} {\bibfnamefont {J.~R.}\ \bibnamefont
  {McClean}}, \bibinfo {author} {\bibfnamefont {J.}~\bibnamefont {Romero}},
  \bibinfo {author} {\bibfnamefont {R.}~\bibnamefont {Babbush}}, \ and\
  \bibinfo {author} {\bibfnamefont {A.}~\bibnamefont {Aspuru-Guzik}},\ }\href
  {\doibase 10.1088/1367-2630/18/2/023023} {\bibfield  {journal} {\bibinfo
  {journal} {New Journal of Physics}\ }\textbf {\bibinfo {volume} {18}},\
  \bibinfo {pages} {023023} (\bibinfo {year} {2016})},\ \bibinfo {note} {arXiv:
  1509.04279}\BibitemShut {NoStop}%
\bibitem [{\citenamefont {McClean}\ \emph {et~al.}(2017)\citenamefont
  {McClean}, \citenamefont {Schwartz}, \citenamefont {Carter},\ and\
  \citenamefont {de~Jong}}]{mcclean_hybrid_2017}%
  \BibitemOpen
  \bibfield  {author} {\bibinfo {author} {\bibfnamefont {J.~R.}\ \bibnamefont
  {McClean}}, \bibinfo {author} {\bibfnamefont {M.~E.}\ \bibnamefont
  {Schwartz}}, \bibinfo {author} {\bibfnamefont {J.}~\bibnamefont {Carter}}, \
  and\ \bibinfo {author} {\bibfnamefont {W.~A.}\ \bibnamefont {de~Jong}},\
  }\href {\doibase 10.1103/PhysRevA.95.042308} {\bibfield  {journal} {\bibinfo
  {journal} {Phys. Rev. A}\ }\textbf {\bibinfo {volume} {95}},\ \bibinfo
  {pages} {042308} (\bibinfo {year} {2017})},\ \bibinfo {note} {arXiv:
  1603.05681}\BibitemShut {NoStop}%
\bibitem [{\citenamefont {Parrish}\ \emph {et~al.}(2019)\citenamefont
  {Parrish}, \citenamefont {Hohenstein}, \citenamefont {McMahon},\ and\
  \citenamefont {Martinez}}]{parrish_quantum_2019}%
  \BibitemOpen
  \bibfield  {author} {\bibinfo {author} {\bibfnamefont {R.~M.}\ \bibnamefont
  {Parrish}}, \bibinfo {author} {\bibfnamefont {E.~G.}\ \bibnamefont
  {Hohenstein}}, \bibinfo {author} {\bibfnamefont {P.~L.}\ \bibnamefont
  {McMahon}}, \ and\ \bibinfo {author} {\bibfnamefont {T.~J.}\ \bibnamefont
  {Martinez}},\ }\href {\doibase 10.1103/PhysRevLett.122.230401} {\bibfield
  {journal} {\bibinfo  {journal} {Phys. Rev. Lett.}\ }\textbf {\bibinfo
  {volume} {122}},\ \bibinfo {pages} {230401} (\bibinfo {year} {2019})},\
  \bibinfo {note} {arXiv: 1901.01234}\BibitemShut {NoStop}%
\bibitem [{\citenamefont {Nakanishi}\ \emph {et~al.}(2019)\citenamefont
  {Nakanishi}, \citenamefont {Mitarai},\ and\ \citenamefont
  {Fujii}}]{nakanishi_subspace-search_2019}%
  \BibitemOpen
  \bibfield  {author} {\bibinfo {author} {\bibfnamefont {K.~M.}\ \bibnamefont
  {Nakanishi}}, \bibinfo {author} {\bibfnamefont {K.}~\bibnamefont {Mitarai}},
  \ and\ \bibinfo {author} {\bibfnamefont {K.}~\bibnamefont {Fujii}},\ }\href
  {http://arxiv.org/abs/1810.09434} {\bibfield  {journal} {\bibinfo  {journal}
  {arXiv:1810.09434 [quant-ph]}\ } (\bibinfo {year} {2019})},\ \bibinfo {note}
  {arXiv: 1810.09434}\BibitemShut {NoStop}%
\bibitem [{\citenamefont {{Sam McArdle}}\ \emph {et~al.}(2018)\citenamefont
  {{Sam McArdle}}, \citenamefont {{Suguru Endo}}, \citenamefont {{Alan
  Aspuru-Guzik}}, \citenamefont {{Simon Benjamin}},\ and\ \citenamefont {{Xiao
  Yuan}}}]{sam_mcardle_quantum_2018}%
  \BibitemOpen
  \bibfield  {author} {\bibinfo {author} {\bibnamefont {{Sam McArdle}}},
  \bibinfo {author} {\bibnamefont {{Suguru Endo}}}, \bibinfo {author}
  {\bibnamefont {{Alan Aspuru-Guzik}}}, \bibinfo {author} {\bibnamefont {{Simon
  Benjamin}}}, \ and\ \bibinfo {author} {\bibnamefont {{Xiao Yuan}}},\ }\href
  {http://arxiv.org/abs/1808.10402} {\bibfield  {journal} {\bibinfo  {journal}
  {arXiv:1808.10402 [quant-ph]}\ } (\bibinfo {year} {2018})},\ \bibinfo {note}
  {arXiv: 1808.10402}\BibitemShut {NoStop}%
\bibitem [{\citenamefont {Jiang}\ \emph {et~al.}(2018)\citenamefont {Jiang},
  \citenamefont {Sung}, \citenamefont {Kechedzhi}, \citenamefont
  {Smelyanskiy},\ and\ \citenamefont {Boixo}}]{jiang_quantum_2018}%
  \BibitemOpen
  \bibfield  {author} {\bibinfo {author} {\bibfnamefont {Z.}~\bibnamefont
  {Jiang}}, \bibinfo {author} {\bibfnamefont {K.~J.}\ \bibnamefont {Sung}},
  \bibinfo {author} {\bibfnamefont {K.}~\bibnamefont {Kechedzhi}}, \bibinfo
  {author} {\bibfnamefont {V.~N.}\ \bibnamefont {Smelyanskiy}}, \ and\ \bibinfo
  {author} {\bibfnamefont {S.}~\bibnamefont {Boixo}},\ }\href {\doibase
  10.1103/PhysRevApplied.9.044036} {\bibfield  {journal} {\bibinfo  {journal}
  {Phys. Rev. Applied}\ }\textbf {\bibinfo {volume} {9}},\ \bibinfo {pages}
  {044036} (\bibinfo {year} {2018})}\BibitemShut {NoStop}%
\bibitem [{\citenamefont {Endo}\ \emph {et~al.}(2020)\citenamefont {Endo},
  \citenamefont {Sun}, \citenamefont {Li}, \citenamefont {Benjamin},\ and\
  \citenamefont {Yuan}}]{endo_variational_2020}%
  \BibitemOpen
  \bibfield  {author} {\bibinfo {author} {\bibfnamefont {S.}~\bibnamefont
  {Endo}}, \bibinfo {author} {\bibfnamefont {J.}~\bibnamefont {Sun}}, \bibinfo
  {author} {\bibfnamefont {Y.}~\bibnamefont {Li}}, \bibinfo {author}
  {\bibfnamefont {S.~C.}\ \bibnamefont {Benjamin}}, \ and\ \bibinfo {author}
  {\bibfnamefont {X.}~\bibnamefont {Yuan}},\ }\href {\doibase
  10.1103/PhysRevLett.125.010501} {\bibfield  {journal} {\bibinfo  {journal}
  {Physical Review Letters}\ }\textbf {\bibinfo {volume} {125}} (\bibinfo
  {year} {2020}),\ 10.1103/PhysRevLett.125.010501}\BibitemShut {NoStop}%
\bibitem [{\citenamefont {Arute}\ \emph {et~al.}(2020)\citenamefont {Arute},
  \citenamefont {Arya}, \citenamefont {Babbush}, \citenamefont {Bacon},
  \citenamefont {Bardin}, \citenamefont {Barends}, \citenamefont {Boixo},
  \citenamefont {Broughton}, \citenamefont {Buckley}, \citenamefont {Buell},
  \citenamefont {Burkett}, \citenamefont {Bushnell}, \citenamefont {Chen},
  \citenamefont {Chen}, \citenamefont {Chiaro}, \citenamefont {Collins},
  \citenamefont {Courtney}, \citenamefont {Demura}, \citenamefont {Dunsworth},
  \citenamefont {Farhi}, \citenamefont {Fowler}, \citenamefont {Foxen},
  \citenamefont {Gidney}, \citenamefont {Giustina}, \citenamefont {Graff},
  \citenamefont {Habegger}, \citenamefont {Harrigan}, \citenamefont {Ho},
  \citenamefont {Hong}, \citenamefont {Huang}, \citenamefont {Huggins},
  \citenamefont {Ioffe}, \citenamefont {Isakov}, \citenamefont {Jeffrey},
  \citenamefont {Jiang}, \citenamefont {Jones}, \citenamefont {Kafri},
  \citenamefont {Kechedzhi}, \citenamefont {Kelly}, \citenamefont {Kim},
  \citenamefont {Klimov}, \citenamefont {Korotkov}, \citenamefont {Kostritsa},
  \citenamefont {Landhuis}, \citenamefont {Laptev}, \citenamefont {Lindmark},
  \citenamefont {Lucero}, \citenamefont {Martin}, \citenamefont {Martinis},
  \citenamefont {McClean}, \citenamefont {McEwen}, \citenamefont {Megrant},
  \citenamefont {Mi}, \citenamefont {Mohseni}, \citenamefont {Mruczkiewicz},
  \citenamefont {Mutus}, \citenamefont {Naaman}, \citenamefont {Neeley},
  \citenamefont {Neill}, \citenamefont {Neven}, \citenamefont {Niu},
  \citenamefont {O'Brien}, \citenamefont {Ostby}, \citenamefont {Petukhov},
  \citenamefont {Putterman}, \citenamefont {Quintana}, \citenamefont {Roushan},
  \citenamefont {Rubin}, \citenamefont {Sank}, \citenamefont {Satzinger},
  \citenamefont {Smelyanskiy}, \citenamefont {Strain}, \citenamefont {Sung},
  \citenamefont {Szalay}, \citenamefont {Takeshita}, \citenamefont
  {Vainsencher}, \citenamefont {White}, \citenamefont {Wiebe}, \citenamefont
  {Yao}, \citenamefont {Yeh},\ and\ \citenamefont
  {Zalcman}}]{arute_hartree-fock_2020}%
  \BibitemOpen
  \bibfield  {author} {\bibinfo {author} {\bibfnamefont {F.}~\bibnamefont
  {Arute}}, \bibinfo {author} {\bibfnamefont {K.}~\bibnamefont {Arya}},
  \bibinfo {author} {\bibfnamefont {R.}~\bibnamefont {Babbush}}, \bibinfo
  {author} {\bibfnamefont {D.}~\bibnamefont {Bacon}}, \bibinfo {author}
  {\bibfnamefont {J.~C.}\ \bibnamefont {Bardin}}, \bibinfo {author}
  {\bibfnamefont {R.}~\bibnamefont {Barends}}, \bibinfo {author} {\bibfnamefont
  {S.}~\bibnamefont {Boixo}}, \bibinfo {author} {\bibfnamefont
  {M.}~\bibnamefont {Broughton}}, \bibinfo {author} {\bibfnamefont {B.~B.}\
  \bibnamefont {Buckley}}, \bibinfo {author} {\bibfnamefont {D.~A.}\
  \bibnamefont {Buell}}, \bibinfo {author} {\bibfnamefont {B.}~\bibnamefont
  {Burkett}}, \bibinfo {author} {\bibfnamefont {N.}~\bibnamefont {Bushnell}},
  \bibinfo {author} {\bibfnamefont {Y.}~\bibnamefont {Chen}}, \bibinfo {author}
  {\bibfnamefont {Z.}~\bibnamefont {Chen}}, \bibinfo {author} {\bibfnamefont
  {B.}~\bibnamefont {Chiaro}}, \bibinfo {author} {\bibfnamefont
  {R.}~\bibnamefont {Collins}}, \bibinfo {author} {\bibfnamefont
  {W.}~\bibnamefont {Courtney}}, \bibinfo {author} {\bibfnamefont
  {S.}~\bibnamefont {Demura}}, \bibinfo {author} {\bibfnamefont
  {A.}~\bibnamefont {Dunsworth}}, \bibinfo {author} {\bibfnamefont
  {E.}~\bibnamefont {Farhi}}, \bibinfo {author} {\bibfnamefont
  {A.}~\bibnamefont {Fowler}}, \bibinfo {author} {\bibfnamefont
  {B.}~\bibnamefont {Foxen}}, \bibinfo {author} {\bibfnamefont
  {C.}~\bibnamefont {Gidney}}, \bibinfo {author} {\bibfnamefont
  {M.}~\bibnamefont {Giustina}}, \bibinfo {author} {\bibfnamefont
  {R.}~\bibnamefont {Graff}}, \bibinfo {author} {\bibfnamefont
  {S.}~\bibnamefont {Habegger}}, \bibinfo {author} {\bibfnamefont {M.~P.}\
  \bibnamefont {Harrigan}}, \bibinfo {author} {\bibfnamefont {A.}~\bibnamefont
  {Ho}}, \bibinfo {author} {\bibfnamefont {S.}~\bibnamefont {Hong}}, \bibinfo
  {author} {\bibfnamefont {T.}~\bibnamefont {Huang}}, \bibinfo {author}
  {\bibfnamefont {W.~J.}\ \bibnamefont {Huggins}}, \bibinfo {author}
  {\bibfnamefont {L.}~\bibnamefont {Ioffe}}, \bibinfo {author} {\bibfnamefont
  {S.~V.}\ \bibnamefont {Isakov}}, \bibinfo {author} {\bibfnamefont
  {E.}~\bibnamefont {Jeffrey}}, \bibinfo {author} {\bibfnamefont
  {Z.}~\bibnamefont {Jiang}}, \bibinfo {author} {\bibfnamefont
  {C.}~\bibnamefont {Jones}}, \bibinfo {author} {\bibfnamefont
  {D.}~\bibnamefont {Kafri}}, \bibinfo {author} {\bibfnamefont
  {K.}~\bibnamefont {Kechedzhi}}, \bibinfo {author} {\bibfnamefont
  {J.}~\bibnamefont {Kelly}}, \bibinfo {author} {\bibfnamefont
  {S.}~\bibnamefont {Kim}}, \bibinfo {author} {\bibfnamefont {P.~V.}\
  \bibnamefont {Klimov}}, \bibinfo {author} {\bibfnamefont {A.}~\bibnamefont
  {Korotkov}}, \bibinfo {author} {\bibfnamefont {F.}~\bibnamefont {Kostritsa}},
  \bibinfo {author} {\bibfnamefont {D.}~\bibnamefont {Landhuis}}, \bibinfo
  {author} {\bibfnamefont {P.}~\bibnamefont {Laptev}}, \bibinfo {author}
  {\bibfnamefont {M.}~\bibnamefont {Lindmark}}, \bibinfo {author}
  {\bibfnamefont {E.}~\bibnamefont {Lucero}}, \bibinfo {author} {\bibfnamefont
  {O.}~\bibnamefont {Martin}}, \bibinfo {author} {\bibfnamefont {J.~M.}\
  \bibnamefont {Martinis}}, \bibinfo {author} {\bibfnamefont {J.~R.}\
  \bibnamefont {McClean}}, \bibinfo {author} {\bibfnamefont {M.}~\bibnamefont
  {McEwen}}, \bibinfo {author} {\bibfnamefont {A.}~\bibnamefont {Megrant}},
  \bibinfo {author} {\bibfnamefont {X.}~\bibnamefont {Mi}}, \bibinfo {author}
  {\bibfnamefont {M.}~\bibnamefont {Mohseni}}, \bibinfo {author} {\bibfnamefont
  {W.}~\bibnamefont {Mruczkiewicz}}, \bibinfo {author} {\bibfnamefont
  {J.}~\bibnamefont {Mutus}}, \bibinfo {author} {\bibfnamefont
  {O.}~\bibnamefont {Naaman}}, \bibinfo {author} {\bibfnamefont
  {M.}~\bibnamefont {Neeley}}, \bibinfo {author} {\bibfnamefont
  {C.}~\bibnamefont {Neill}}, \bibinfo {author} {\bibfnamefont
  {H.}~\bibnamefont {Neven}}, \bibinfo {author} {\bibfnamefont {M.~Y.}\
  \bibnamefont {Niu}}, \bibinfo {author} {\bibfnamefont {T.~E.}\ \bibnamefont
  {O'Brien}}, \bibinfo {author} {\bibfnamefont {E.}~\bibnamefont {Ostby}},
  \bibinfo {author} {\bibfnamefont {A.}~\bibnamefont {Petukhov}}, \bibinfo
  {author} {\bibfnamefont {H.}~\bibnamefont {Putterman}}, \bibinfo {author}
  {\bibfnamefont {C.}~\bibnamefont {Quintana}}, \bibinfo {author}
  {\bibfnamefont {P.}~\bibnamefont {Roushan}}, \bibinfo {author} {\bibfnamefont
  {N.~C.}\ \bibnamefont {Rubin}}, \bibinfo {author} {\bibfnamefont
  {D.}~\bibnamefont {Sank}}, \bibinfo {author} {\bibfnamefont {K.~J.}\
  \bibnamefont {Satzinger}}, \bibinfo {author} {\bibfnamefont {V.}~\bibnamefont
  {Smelyanskiy}}, \bibinfo {author} {\bibfnamefont {D.}~\bibnamefont {Strain}},
  \bibinfo {author} {\bibfnamefont {K.~J.}\ \bibnamefont {Sung}}, \bibinfo
  {author} {\bibfnamefont {M.}~\bibnamefont {Szalay}}, \bibinfo {author}
  {\bibfnamefont {T.~Y.}\ \bibnamefont {Takeshita}}, \bibinfo {author}
  {\bibfnamefont {A.}~\bibnamefont {Vainsencher}}, \bibinfo {author}
  {\bibfnamefont {T.}~\bibnamefont {White}}, \bibinfo {author} {\bibfnamefont
  {N.}~\bibnamefont {Wiebe}}, \bibinfo {author} {\bibfnamefont {Z.~J.}\
  \bibnamefont {Yao}}, \bibinfo {author} {\bibfnamefont {P.}~\bibnamefont
  {Yeh}}, \ and\ \bibinfo {author} {\bibfnamefont {A.}~\bibnamefont
  {Zalcman}},\ }\href {http://arxiv.org/abs/2004.04174} {\bibfield  {journal}
  {\bibinfo  {journal} {arXiv:2004.04174 [physics, physics:quant-ph]}\ }
  (\bibinfo {year} {2020})},\ \bibinfo {note} {arXiv: 2004.04174}\BibitemShut
  {NoStop}%
\bibitem [{\citenamefont {Jones}\ \emph {et~al.}(2019)\citenamefont {Jones},
  \citenamefont {Brown}, \citenamefont {Bush},\ and\ \citenamefont
  {Benjamin}}]{jones_quest_2019}%
  \BibitemOpen
  \bibfield  {author} {\bibinfo {author} {\bibfnamefont {T.}~\bibnamefont
  {Jones}}, \bibinfo {author} {\bibfnamefont {A.}~\bibnamefont {Brown}},
  \bibinfo {author} {\bibfnamefont {I.}~\bibnamefont {Bush}}, \ and\ \bibinfo
  {author} {\bibfnamefont {S.~C.}\ \bibnamefont {Benjamin}},\ }\href@noop {}
  {\bibfield  {journal} {\bibinfo  {journal} {Scientific reports}\ }\textbf
  {\bibinfo {volume} {9}},\ \bibinfo {pages} {1} (\bibinfo {year} {2019})},\
  \bibinfo {note} {publisher: Nature Publishing Group}\BibitemShut {NoStop}%
\bibitem [{\citenamefont {Reiner}\ \emph {et~al.}(2019)\citenamefont {Reiner},
  \citenamefont {Wilhelm-Mauch}, \citenamefont {Schön},\ and\ \citenamefont
  {Marthaler}}]{reiner_finding_2018}%
  \BibitemOpen
  \bibfield  {author} {\bibinfo {author} {\bibfnamefont {J.-M.}\ \bibnamefont
  {Reiner}}, \bibinfo {author} {\bibfnamefont {F.}~\bibnamefont
  {Wilhelm-Mauch}}, \bibinfo {author} {\bibfnamefont {G.}~\bibnamefont
  {Schön}}, \ and\ \bibinfo {author} {\bibfnamefont {M.}~\bibnamefont
  {Marthaler}},\ }\href {\doibase 10.1088/2058-9565/ab1e85} {\bibfield
  {journal} {\bibinfo  {journal} {Quantum Science and Technology}\ }\textbf
  {\bibinfo {volume} {4}},\ \bibinfo {pages} {035005} (\bibinfo {year}
  {2019})}\BibitemShut {NoStop}%
\bibitem [{\citenamefont {Nakamura}\ \emph {et~al.}(1999)\citenamefont
  {Nakamura}, \citenamefont {Pashkin},\ and\ \citenamefont
  {Tsai}}]{nakamura_coherent_1999}%
  \BibitemOpen
  \bibfield  {author} {\bibinfo {author} {\bibfnamefont {Y.}~\bibnamefont
  {Nakamura}}, \bibinfo {author} {\bibfnamefont {Y.~A.}\ \bibnamefont
  {Pashkin}}, \ and\ \bibinfo {author} {\bibfnamefont {J.~S.}\ \bibnamefont
  {Tsai}},\ }\href {\doibase 10.1038/19718} {\bibfield  {journal} {\bibinfo
  {journal} {Nature}\ }\textbf {\bibinfo {volume} {398}},\ \bibinfo {pages}
  {786} (\bibinfo {year} {1999})},\ \bibinfo {note} {arXiv:
  cond-mat/9904003}\BibitemShut {NoStop}%
\bibitem [{\citenamefont {Clarke}\ and\ \citenamefont
  {Wilhelm}(2008)}]{clarke_superconducting_2008}%
  \BibitemOpen
  \bibfield  {author} {\bibinfo {author} {\bibfnamefont {J.}~\bibnamefont
  {Clarke}}\ and\ \bibinfo {author} {\bibfnamefont {F.~K.}\ \bibnamefont
  {Wilhelm}},\ }\href {\doibase 10.1038/nature07128} {\bibfield  {journal}
  {\bibinfo  {journal} {Nature}\ }\textbf {\bibinfo {volume} {453}},\ \bibinfo
  {pages} {1031} (\bibinfo {year} {2008})}\BibitemShut {NoStop}%
\bibitem [{\citenamefont {Zwanenburg}\ \emph {et~al.}(2013)\citenamefont
  {Zwanenburg}, \citenamefont {Dzurak}, \citenamefont {Morello}, \citenamefont
  {Simmons}, \citenamefont {Hollenberg}, \citenamefont {Klimeck}, \citenamefont
  {Rogge}, \citenamefont {Coppersmith},\ and\ \citenamefont
  {Eriksson}}]{zwanenburg2013silicon}%
  \BibitemOpen
  \bibfield  {author} {\bibinfo {author} {\bibfnamefont {F.~A.}\ \bibnamefont
  {Zwanenburg}}, \bibinfo {author} {\bibfnamefont {A.~S.}\ \bibnamefont
  {Dzurak}}, \bibinfo {author} {\bibfnamefont {A.}~\bibnamefont {Morello}},
  \bibinfo {author} {\bibfnamefont {M.~Y.}\ \bibnamefont {Simmons}}, \bibinfo
  {author} {\bibfnamefont {L.~C.}\ \bibnamefont {Hollenberg}}, \bibinfo
  {author} {\bibfnamefont {G.}~\bibnamefont {Klimeck}}, \bibinfo {author}
  {\bibfnamefont {S.}~\bibnamefont {Rogge}}, \bibinfo {author} {\bibfnamefont
  {S.~N.}\ \bibnamefont {Coppersmith}}, \ and\ \bibinfo {author} {\bibfnamefont
  {M.~A.}\ \bibnamefont {Eriksson}},\ }\href@noop {} {\bibfield  {journal}
  {\bibinfo  {journal} {Reviews of modern physics}\ }\textbf {\bibinfo {volume}
  {85}},\ \bibinfo {pages} {961} (\bibinfo {year} {2013})}\BibitemShut
  {NoStop}%
\bibitem [{\citenamefont {Veldhorst}\ \emph {et~al.}(2014)\citenamefont
  {Veldhorst}, \citenamefont {Hwang}, \citenamefont {Yang}, \citenamefont
  {Leenstra}, \citenamefont {de~Ronde}, \citenamefont {Dehollain},
  \citenamefont {Muhonen}, \citenamefont {Hudson}, \citenamefont {Itoh},
  \citenamefont {Morello} \emph {et~al.}}]{veldhorst2014addressable}%
  \BibitemOpen
  \bibfield  {author} {\bibinfo {author} {\bibfnamefont {M.}~\bibnamefont
  {Veldhorst}}, \bibinfo {author} {\bibfnamefont {J.}~\bibnamefont {Hwang}},
  \bibinfo {author} {\bibfnamefont {C.}~\bibnamefont {Yang}}, \bibinfo {author}
  {\bibfnamefont {A.}~\bibnamefont {Leenstra}}, \bibinfo {author}
  {\bibfnamefont {B.}~\bibnamefont {de~Ronde}}, \bibinfo {author}
  {\bibfnamefont {J.}~\bibnamefont {Dehollain}}, \bibinfo {author}
  {\bibfnamefont {J.}~\bibnamefont {Muhonen}}, \bibinfo {author} {\bibfnamefont
  {F.}~\bibnamefont {Hudson}}, \bibinfo {author} {\bibfnamefont {K.~M.}\
  \bibnamefont {Itoh}}, \bibinfo {author} {\bibfnamefont {A.}~\bibnamefont
  {Morello}},  \emph {et~al.},\ }\href@noop {} {\bibfield  {journal} {\bibinfo
  {journal} {Nature nanotechnology}\ }\textbf {\bibinfo {volume} {9}},\
  \bibinfo {pages} {981} (\bibinfo {year} {2014})}\BibitemShut {NoStop}%
\bibitem [{\citenamefont {Watson}\ \emph {et~al.}(2018)\citenamefont {Watson},
  \citenamefont {Philips}, \citenamefont {Kawakami}, \citenamefont {Ward},
  \citenamefont {Scarlino}, \citenamefont {Veldhorst}, \citenamefont {Savage},
  \citenamefont {Lagally}, \citenamefont {Friesen}, \citenamefont {Coppersmith}
  \emph {et~al.}}]{watson2018programmable}%
  \BibitemOpen
  \bibfield  {author} {\bibinfo {author} {\bibfnamefont {T.}~\bibnamefont
  {Watson}}, \bibinfo {author} {\bibfnamefont {S.}~\bibnamefont {Philips}},
  \bibinfo {author} {\bibfnamefont {E.}~\bibnamefont {Kawakami}}, \bibinfo
  {author} {\bibfnamefont {D.}~\bibnamefont {Ward}}, \bibinfo {author}
  {\bibfnamefont {P.}~\bibnamefont {Scarlino}}, \bibinfo {author}
  {\bibfnamefont {M.}~\bibnamefont {Veldhorst}}, \bibinfo {author}
  {\bibfnamefont {D.}~\bibnamefont {Savage}}, \bibinfo {author} {\bibfnamefont
  {M.}~\bibnamefont {Lagally}}, \bibinfo {author} {\bibfnamefont
  {M.}~\bibnamefont {Friesen}}, \bibinfo {author} {\bibfnamefont
  {S.}~\bibnamefont {Coppersmith}},  \emph {et~al.},\ }\href@noop {} {\bibfield
   {journal} {\bibinfo  {journal} {Nature}\ }\textbf {\bibinfo {volume}
  {555}},\ \bibinfo {pages} {633} (\bibinfo {year} {2018})}\BibitemShut
  {NoStop}%
\bibitem [{\citenamefont {He}\ \emph {et~al.}(2019)\citenamefont {He},
  \citenamefont {Gorman}, \citenamefont {Keith}, \citenamefont {Kranz},
  \citenamefont {Keizer},\ and\ \citenamefont {Simmons}}]{he2019two}%
  \BibitemOpen
  \bibfield  {author} {\bibinfo {author} {\bibfnamefont {Y.}~\bibnamefont
  {He}}, \bibinfo {author} {\bibfnamefont {S.}~\bibnamefont {Gorman}}, \bibinfo
  {author} {\bibfnamefont {D.}~\bibnamefont {Keith}}, \bibinfo {author}
  {\bibfnamefont {L.}~\bibnamefont {Kranz}}, \bibinfo {author} {\bibfnamefont
  {J.}~\bibnamefont {Keizer}}, \ and\ \bibinfo {author} {\bibfnamefont
  {M.}~\bibnamefont {Simmons}},\ }\href@noop {} {\bibfield  {journal} {\bibinfo
   {journal} {Nature}\ }\textbf {\bibinfo {volume} {571}},\ \bibinfo {pages}
  {371} (\bibinfo {year} {2019})}\BibitemShut {NoStop}%
\bibitem [{\citenamefont {Cirac}\ and\ \citenamefont
  {Zoller}(1995)}]{cirac_quantum_1995}%
  \BibitemOpen
  \bibfield  {author} {\bibinfo {author} {\bibfnamefont {J.~I.}\ \bibnamefont
  {Cirac}}\ and\ \bibinfo {author} {\bibfnamefont {P.}~\bibnamefont {Zoller}},\
  }\href {\doibase 10.1103/PhysRevLett.74.4091} {\bibfield  {journal} {\bibinfo
   {journal} {Physical Review Letters}\ }\textbf {\bibinfo {volume} {74}},\
  \bibinfo {pages} {4091} (\bibinfo {year} {1995})}\BibitemShut {NoStop}%
\bibitem [{\citenamefont {Bermudez}\ \emph {et~al.}(2017)\citenamefont
  {Bermudez}, \citenamefont {Xu}, \citenamefont {Nigmatullin}, \citenamefont
  {O'Gorman}, \citenamefont {Negnevitsky}, \citenamefont {Schindler},
  \citenamefont {Monz}, \citenamefont {Poschinger}, \citenamefont {Hempel},
  \citenamefont {Home}, \citenamefont {Schmidt-Kaler}, \citenamefont {Biercuk},
  \citenamefont {Blatt}, \citenamefont {Benjamin},\ and\ \citenamefont
  {Müller}}]{bermudez_assessing_2017}%
  \BibitemOpen
  \bibfield  {author} {\bibinfo {author} {\bibfnamefont {A.}~\bibnamefont
  {Bermudez}}, \bibinfo {author} {\bibfnamefont {X.}~\bibnamefont {Xu}},
  \bibinfo {author} {\bibfnamefont {R.}~\bibnamefont {Nigmatullin}}, \bibinfo
  {author} {\bibfnamefont {J.}~\bibnamefont {O'Gorman}}, \bibinfo {author}
  {\bibfnamefont {V.}~\bibnamefont {Negnevitsky}}, \bibinfo {author}
  {\bibfnamefont {P.}~\bibnamefont {Schindler}}, \bibinfo {author}
  {\bibfnamefont {T.}~\bibnamefont {Monz}}, \bibinfo {author} {\bibfnamefont
  {U.~G.}\ \bibnamefont {Poschinger}}, \bibinfo {author} {\bibfnamefont
  {C.}~\bibnamefont {Hempel}}, \bibinfo {author} {\bibfnamefont
  {J.}~\bibnamefont {Home}}, \bibinfo {author} {\bibfnamefont {F.}~\bibnamefont
  {Schmidt-Kaler}}, \bibinfo {author} {\bibfnamefont {M.}~\bibnamefont
  {Biercuk}}, \bibinfo {author} {\bibfnamefont {R.}~\bibnamefont {Blatt}},
  \bibinfo {author} {\bibfnamefont {S.}~\bibnamefont {Benjamin}}, \ and\
  \bibinfo {author} {\bibfnamefont {M.}~\bibnamefont {Müller}},\ }\href
  {\doibase 10.1103/PhysRevX.7.041061} {\bibfield  {journal} {\bibinfo
  {journal} {Phys. Rev. X}\ }\textbf {\bibinfo {volume} {7}},\ \bibinfo {pages}
  {041061} (\bibinfo {year} {2017})},\ \bibinfo {note} {arXiv:
  1705.02771}\BibitemShut {NoStop}%
\bibitem [{\citenamefont {Kivlichan}\ \emph {et~al.}(2018)\citenamefont
  {Kivlichan}, \citenamefont {McClean}, \citenamefont {Wiebe}, \citenamefont
  {Gidney}, \citenamefont {Aspuru-Guzik}, \citenamefont {Chan},\ and\
  \citenamefont {Babbush}}]{kivlichan_quantum_2018}%
  \BibitemOpen
  \bibfield  {author} {\bibinfo {author} {\bibfnamefont {I.~D.}\ \bibnamefont
  {Kivlichan}}, \bibinfo {author} {\bibfnamefont {J.}~\bibnamefont {McClean}},
  \bibinfo {author} {\bibfnamefont {N.}~\bibnamefont {Wiebe}}, \bibinfo
  {author} {\bibfnamefont {C.}~\bibnamefont {Gidney}}, \bibinfo {author}
  {\bibfnamefont {A.}~\bibnamefont {Aspuru-Guzik}}, \bibinfo {author}
  {\bibfnamefont {G.~K.-L.}\ \bibnamefont {Chan}}, \ and\ \bibinfo {author}
  {\bibfnamefont {R.}~\bibnamefont {Babbush}},\ }\href {\doibase
  10.1103/PhysRevLett.120.110501} {\bibfield  {journal} {\bibinfo  {journal}
  {Physical Review Letters}\ }\textbf {\bibinfo {volume} {120}} (\bibinfo
  {year} {2018}),\ 10.1103/PhysRevLett.120.110501},\ \bibinfo {note} {arXiv:
  1711.04789}\BibitemShut {NoStop}%
\bibitem [{\citenamefont {McClean}\ \emph {et~al.}(2019)\citenamefont
  {McClean}, \citenamefont {Sung}, \citenamefont {Kivlichan}, \citenamefont
  {Cao}, \citenamefont {Dai}, \citenamefont {Fried}, \citenamefont {Gidney},
  \citenamefont {Gimby}, \citenamefont {Gokhale}, \citenamefont {Häner},
  \citenamefont {Hardikar}, \citenamefont {Havlíček}, \citenamefont
  {Higgott}, \citenamefont {Huang}, \citenamefont {Izaac}, \citenamefont
  {Jiang}, \citenamefont {Liu}, \citenamefont {McArdle}, \citenamefont
  {Neeley}, \citenamefont {O'Brien}, \citenamefont {O'Gorman}, \citenamefont
  {Ozfidan}, \citenamefont {Radin}, \citenamefont {Romero}, \citenamefont
  {Rubin}, \citenamefont {Sawaya}, \citenamefont {Setia}, \citenamefont {Sim},
  \citenamefont {Steiger}, \citenamefont {Steudtner}, \citenamefont {Sun},
  \citenamefont {Sun}, \citenamefont {Wang}, \citenamefont {Zhang},\ and\
  \citenamefont {Babbush}}]{mcclean_openfermion_2019}%
  \BibitemOpen
  \bibfield  {author} {\bibinfo {author} {\bibfnamefont {J.~R.}\ \bibnamefont
  {McClean}}, \bibinfo {author} {\bibfnamefont {K.~J.}\ \bibnamefont {Sung}},
  \bibinfo {author} {\bibfnamefont {I.~D.}\ \bibnamefont {Kivlichan}}, \bibinfo
  {author} {\bibfnamefont {Y.}~\bibnamefont {Cao}}, \bibinfo {author}
  {\bibfnamefont {C.}~\bibnamefont {Dai}}, \bibinfo {author} {\bibfnamefont
  {E.~S.}\ \bibnamefont {Fried}}, \bibinfo {author} {\bibfnamefont
  {C.}~\bibnamefont {Gidney}}, \bibinfo {author} {\bibfnamefont
  {B.}~\bibnamefont {Gimby}}, \bibinfo {author} {\bibfnamefont
  {P.}~\bibnamefont {Gokhale}}, \bibinfo {author} {\bibfnamefont
  {T.}~\bibnamefont {Häner}}, \bibinfo {author} {\bibfnamefont
  {T.}~\bibnamefont {Hardikar}}, \bibinfo {author} {\bibfnamefont
  {V.}~\bibnamefont {Havlíček}}, \bibinfo {author} {\bibfnamefont
  {O.}~\bibnamefont {Higgott}}, \bibinfo {author} {\bibfnamefont
  {C.}~\bibnamefont {Huang}}, \bibinfo {author} {\bibfnamefont
  {J.}~\bibnamefont {Izaac}}, \bibinfo {author} {\bibfnamefont
  {Z.}~\bibnamefont {Jiang}}, \bibinfo {author} {\bibfnamefont
  {X.}~\bibnamefont {Liu}}, \bibinfo {author} {\bibfnamefont {S.}~\bibnamefont
  {McArdle}}, \bibinfo {author} {\bibfnamefont {M.}~\bibnamefont {Neeley}},
  \bibinfo {author} {\bibfnamefont {T.}~\bibnamefont {O'Brien}}, \bibinfo
  {author} {\bibfnamefont {B.}~\bibnamefont {O'Gorman}}, \bibinfo {author}
  {\bibfnamefont {I.}~\bibnamefont {Ozfidan}}, \bibinfo {author} {\bibfnamefont
  {M.~D.}\ \bibnamefont {Radin}}, \bibinfo {author} {\bibfnamefont
  {J.}~\bibnamefont {Romero}}, \bibinfo {author} {\bibfnamefont
  {N.}~\bibnamefont {Rubin}}, \bibinfo {author} {\bibfnamefont {N.~P.~D.}\
  \bibnamefont {Sawaya}}, \bibinfo {author} {\bibfnamefont {K.}~\bibnamefont
  {Setia}}, \bibinfo {author} {\bibfnamefont {S.}~\bibnamefont {Sim}}, \bibinfo
  {author} {\bibfnamefont {D.~S.}\ \bibnamefont {Steiger}}, \bibinfo {author}
  {\bibfnamefont {M.}~\bibnamefont {Steudtner}}, \bibinfo {author}
  {\bibfnamefont {Q.}~\bibnamefont {Sun}}, \bibinfo {author} {\bibfnamefont
  {W.}~\bibnamefont {Sun}}, \bibinfo {author} {\bibfnamefont {D.}~\bibnamefont
  {Wang}}, \bibinfo {author} {\bibfnamefont {F.}~\bibnamefont {Zhang}}, \ and\
  \bibinfo {author} {\bibfnamefont {R.}~\bibnamefont {Babbush}},\ }\href
  {http://arxiv.org/abs/1710.07629} {\bibfield  {journal} {\bibinfo  {journal}
  {arXiv:1710.07629 [physics, physics:quant-ph]}\ } (\bibinfo {year} {2019})},\
  \bibinfo {note} {arXiv: 1710.07629}\BibitemShut {NoStop}%
\bibitem [{\citenamefont {Zhu}\ \emph {et~al.}(1997)\citenamefont {Zhu},
  \citenamefont {Byrd}, \citenamefont {Lu},\ and\ \citenamefont
  {Nocedal}}]{lbfgsb}%
  \BibitemOpen
  \bibfield  {author} {\bibinfo {author} {\bibfnamefont {C.}~\bibnamefont
  {Zhu}}, \bibinfo {author} {\bibfnamefont {R.~H.}\ \bibnamefont {Byrd}},
  \bibinfo {author} {\bibfnamefont {P.}~\bibnamefont {Lu}}, \ and\ \bibinfo
  {author} {\bibfnamefont {J.}~\bibnamefont {Nocedal}},\ }\href@noop {}
  {\bibfield  {journal} {\bibinfo  {journal} {ACM Trans. Math. Softw.}\
  }\textbf {\bibinfo {volume} {23}},\ \bibinfo {pages} {550} (\bibinfo {year}
  {1997})}\BibitemShut {NoStop}%
\bibitem [{\citenamefont {Temme}\ \emph {et~al.}(2017)\citenamefont {Temme},
  \citenamefont {Bravyi},\ and\ \citenamefont
  {Gambetta}}]{Temme_extrapolation_17}%
  \BibitemOpen
  \bibfield  {author} {\bibinfo {author} {\bibfnamefont {K.}~\bibnamefont
  {Temme}}, \bibinfo {author} {\bibfnamefont {S.}~\bibnamefont {Bravyi}}, \
  and\ \bibinfo {author} {\bibfnamefont {J.~M.}\ \bibnamefont {Gambetta}},\
  }\href {\doibase 10.1103/PhysRevLett.119.180509} {\bibfield  {journal}
  {\bibinfo  {journal} {Physical Review Letters}\ }\textbf {\bibinfo {volume}
  {119}},\ \bibinfo {pages} {180509} (\bibinfo {year} {2017})}\BibitemShut
  {NoStop}%
\bibitem [{\citenamefont {Arute}\ \emph {et~al.}(2019)\citenamefont {Arute},
  \citenamefont {Arya}, \citenamefont {Babbush}, \citenamefont {Bacon},
  \citenamefont {Bardin}, \citenamefont {Barends}, \citenamefont {Biswas},
  \citenamefont {Boixo}, \citenamefont {Brandao}, \citenamefont {Buell},
  \citenamefont {Burkett}, \citenamefont {Chen}, \citenamefont {Chen},
  \citenamefont {Chiaro}, \citenamefont {Collins}, \citenamefont {Courtney},
  \citenamefont {Dunsworth}, \citenamefont {Farhi}, \citenamefont {Foxen},
  \citenamefont {Fowler}, \citenamefont {Gidney}, \citenamefont {Giustina},
  \citenamefont {Graff}, \citenamefont {Guerin}, \citenamefont {Habegger},
  \citenamefont {Harrigan}, \citenamefont {Hartmann}, \citenamefont {Ho},
  \citenamefont {Hoffmann}, \citenamefont {Huang}, \citenamefont {Humble},
  \citenamefont {Isakov}, \citenamefont {Jeffrey}, \citenamefont {Jiang},
  \citenamefont {Kafri}, \citenamefont {Kechedzhi}, \citenamefont {Kelly},
  \citenamefont {Klimov}, \citenamefont {Knysh}, \citenamefont {Korotkov},
  \citenamefont {Kostritsa}, \citenamefont {Landhuis}, \citenamefont
  {Lindmark}, \citenamefont {Lucero}, \citenamefont {Lyakh}, \citenamefont
  {Mandrà}, \citenamefont {McClean}, \citenamefont {McEwen}, \citenamefont
  {Megrant}, \citenamefont {Mi}, \citenamefont {Michielsen}, \citenamefont
  {Mohseni}, \citenamefont {Mutus}, \citenamefont {Naaman}, \citenamefont
  {Neeley}, \citenamefont {Neill}, \citenamefont {Niu}, \citenamefont {Ostby},
  \citenamefont {Petukhov}, \citenamefont {Platt}, \citenamefont {Quintana},
  \citenamefont {Rieffel}, \citenamefont {Roushan}, \citenamefont {Rubin},
  \citenamefont {Sank}, \citenamefont {Satzinger}, \citenamefont {Smelyanskiy},
  \citenamefont {Sung}, \citenamefont {Trevithick}, \citenamefont
  {Vainsencher}, \citenamefont {Villalonga}, \citenamefont {White},
  \citenamefont {Yao}, \citenamefont {Yeh}, \citenamefont {Zalcman},
  \citenamefont {Neven},\ and\ \citenamefont {Martinis}}]{arute_quantum_2019}%
  \BibitemOpen
  \bibfield  {author} {\bibinfo {author} {\bibfnamefont {F.}~\bibnamefont
  {Arute}}, \bibinfo {author} {\bibfnamefont {K.}~\bibnamefont {Arya}},
  \bibinfo {author} {\bibfnamefont {R.}~\bibnamefont {Babbush}}, \bibinfo
  {author} {\bibfnamefont {D.}~\bibnamefont {Bacon}}, \bibinfo {author}
  {\bibfnamefont {J.~C.}\ \bibnamefont {Bardin}}, \bibinfo {author}
  {\bibfnamefont {R.}~\bibnamefont {Barends}}, \bibinfo {author} {\bibfnamefont
  {R.}~\bibnamefont {Biswas}}, \bibinfo {author} {\bibfnamefont
  {S.}~\bibnamefont {Boixo}}, \bibinfo {author} {\bibfnamefont {F.~G. S.~L.}\
  \bibnamefont {Brandao}}, \bibinfo {author} {\bibfnamefont {D.~A.}\
  \bibnamefont {Buell}}, \bibinfo {author} {\bibfnamefont {B.}~\bibnamefont
  {Burkett}}, \bibinfo {author} {\bibfnamefont {Y.}~\bibnamefont {Chen}},
  \bibinfo {author} {\bibfnamefont {Z.}~\bibnamefont {Chen}}, \bibinfo {author}
  {\bibfnamefont {B.}~\bibnamefont {Chiaro}}, \bibinfo {author} {\bibfnamefont
  {R.}~\bibnamefont {Collins}}, \bibinfo {author} {\bibfnamefont
  {W.}~\bibnamefont {Courtney}}, \bibinfo {author} {\bibfnamefont
  {A.}~\bibnamefont {Dunsworth}}, \bibinfo {author} {\bibfnamefont
  {E.}~\bibnamefont {Farhi}}, \bibinfo {author} {\bibfnamefont
  {B.}~\bibnamefont {Foxen}}, \bibinfo {author} {\bibfnamefont
  {A.}~\bibnamefont {Fowler}}, \bibinfo {author} {\bibfnamefont
  {C.}~\bibnamefont {Gidney}}, \bibinfo {author} {\bibfnamefont
  {M.}~\bibnamefont {Giustina}}, \bibinfo {author} {\bibfnamefont
  {R.}~\bibnamefont {Graff}}, \bibinfo {author} {\bibfnamefont
  {K.}~\bibnamefont {Guerin}}, \bibinfo {author} {\bibfnamefont
  {S.}~\bibnamefont {Habegger}}, \bibinfo {author} {\bibfnamefont {M.~P.}\
  \bibnamefont {Harrigan}}, \bibinfo {author} {\bibfnamefont {M.~J.}\
  \bibnamefont {Hartmann}}, \bibinfo {author} {\bibfnamefont {A.}~\bibnamefont
  {Ho}}, \bibinfo {author} {\bibfnamefont {M.}~\bibnamefont {Hoffmann}},
  \bibinfo {author} {\bibfnamefont {T.}~\bibnamefont {Huang}}, \bibinfo
  {author} {\bibfnamefont {T.~S.}\ \bibnamefont {Humble}}, \bibinfo {author}
  {\bibfnamefont {S.~V.}\ \bibnamefont {Isakov}}, \bibinfo {author}
  {\bibfnamefont {E.}~\bibnamefont {Jeffrey}}, \bibinfo {author} {\bibfnamefont
  {Z.}~\bibnamefont {Jiang}}, \bibinfo {author} {\bibfnamefont
  {D.}~\bibnamefont {Kafri}}, \bibinfo {author} {\bibfnamefont
  {K.}~\bibnamefont {Kechedzhi}}, \bibinfo {author} {\bibfnamefont
  {J.}~\bibnamefont {Kelly}}, \bibinfo {author} {\bibfnamefont {P.~V.}\
  \bibnamefont {Klimov}}, \bibinfo {author} {\bibfnamefont {S.}~\bibnamefont
  {Knysh}}, \bibinfo {author} {\bibfnamefont {A.}~\bibnamefont {Korotkov}},
  \bibinfo {author} {\bibfnamefont {F.}~\bibnamefont {Kostritsa}}, \bibinfo
  {author} {\bibfnamefont {D.}~\bibnamefont {Landhuis}}, \bibinfo {author}
  {\bibfnamefont {M.}~\bibnamefont {Lindmark}}, \bibinfo {author}
  {\bibfnamefont {E.}~\bibnamefont {Lucero}}, \bibinfo {author} {\bibfnamefont
  {D.}~\bibnamefont {Lyakh}}, \bibinfo {author} {\bibfnamefont
  {S.}~\bibnamefont {Mandrà}}, \bibinfo {author} {\bibfnamefont {J.~R.}\
  \bibnamefont {McClean}}, \bibinfo {author} {\bibfnamefont {M.}~\bibnamefont
  {McEwen}}, \bibinfo {author} {\bibfnamefont {A.}~\bibnamefont {Megrant}},
  \bibinfo {author} {\bibfnamefont {X.}~\bibnamefont {Mi}}, \bibinfo {author}
  {\bibfnamefont {K.}~\bibnamefont {Michielsen}}, \bibinfo {author}
  {\bibfnamefont {M.}~\bibnamefont {Mohseni}}, \bibinfo {author} {\bibfnamefont
  {J.}~\bibnamefont {Mutus}}, \bibinfo {author} {\bibfnamefont
  {O.}~\bibnamefont {Naaman}}, \bibinfo {author} {\bibfnamefont
  {M.}~\bibnamefont {Neeley}}, \bibinfo {author} {\bibfnamefont
  {C.}~\bibnamefont {Neill}}, \bibinfo {author} {\bibfnamefont {M.~Y.}\
  \bibnamefont {Niu}}, \bibinfo {author} {\bibfnamefont {E.}~\bibnamefont
  {Ostby}}, \bibinfo {author} {\bibfnamefont {A.}~\bibnamefont {Petukhov}},
  \bibinfo {author} {\bibfnamefont {J.~C.}\ \bibnamefont {Platt}}, \bibinfo
  {author} {\bibfnamefont {C.}~\bibnamefont {Quintana}}, \bibinfo {author}
  {\bibfnamefont {E.~G.}\ \bibnamefont {Rieffel}}, \bibinfo {author}
  {\bibfnamefont {P.}~\bibnamefont {Roushan}}, \bibinfo {author} {\bibfnamefont
  {N.~C.}\ \bibnamefont {Rubin}}, \bibinfo {author} {\bibfnamefont
  {D.}~\bibnamefont {Sank}}, \bibinfo {author} {\bibfnamefont {K.~J.}\
  \bibnamefont {Satzinger}}, \bibinfo {author} {\bibfnamefont {V.}~\bibnamefont
  {Smelyanskiy}}, \bibinfo {author} {\bibfnamefont {K.~J.}\ \bibnamefont
  {Sung}}, \bibinfo {author} {\bibfnamefont {M.~D.}\ \bibnamefont
  {Trevithick}}, \bibinfo {author} {\bibfnamefont {A.}~\bibnamefont
  {Vainsencher}}, \bibinfo {author} {\bibfnamefont {B.}~\bibnamefont
  {Villalonga}}, \bibinfo {author} {\bibfnamefont {T.}~\bibnamefont {White}},
  \bibinfo {author} {\bibfnamefont {Z.~J.}\ \bibnamefont {Yao}}, \bibinfo
  {author} {\bibfnamefont {P.}~\bibnamefont {Yeh}}, \bibinfo {author}
  {\bibfnamefont {A.}~\bibnamefont {Zalcman}}, \bibinfo {author} {\bibfnamefont
  {H.}~\bibnamefont {Neven}}, \ and\ \bibinfo {author} {\bibfnamefont {J.~M.}\
  \bibnamefont {Martinis}},\ }\href {\doibase 10.1038/s41586-019-1666-5}
  {\bibfield  {journal} {\bibinfo  {journal} {Nature}\ }\textbf {\bibinfo
  {volume} {574}},\ \bibinfo {pages} {505} (\bibinfo {year}
  {2019})}\BibitemShut {NoStop}%
\end{thebibliography}%
\appendix
\section{Stochastic Measurement}
On a real quantum computer the simplest way to measure the Hubbard Hamiltonian \(H\) is to rewrite it as a sum over Pauli products with the Jordan-Wigner Transform,
\begin{align}
  H &= \sum_{\gamma} q_{\gamma} \prod_{i}^{M} \sigma_{i,\gamma} \\
  \sigma_{i,\gamma} &\in \left\{I_I, \sigma^x_i, \sigma^y_i, \sigma^z_I \right\} \ ,
\end{align}
where the index \(i\) runs over all \(M\) orbitals and the operators \(\sigma_{i,\gamma}\) are the identity or Pauli operators acting on the \(i\)-th qubit.
The real numbers \(q_{\gamma}\) are the prefactors of the Pauli products in the Hamiltonian.
Each product of Pauli operators can be measured on the quantum computer by rotating each qubit into the correct basis and performing a projective measurement in the physical \(\sigma^z\) basis of the qubit.
Since each projective measurement of a Pauli product returns either \(-1\) or \(1\), a large number of measurements is necessary to obtain the expectation value in the interval \([-1,1]\).

\begin{figure}
  \centering
  \includegraphics[width=.5\textwidth]{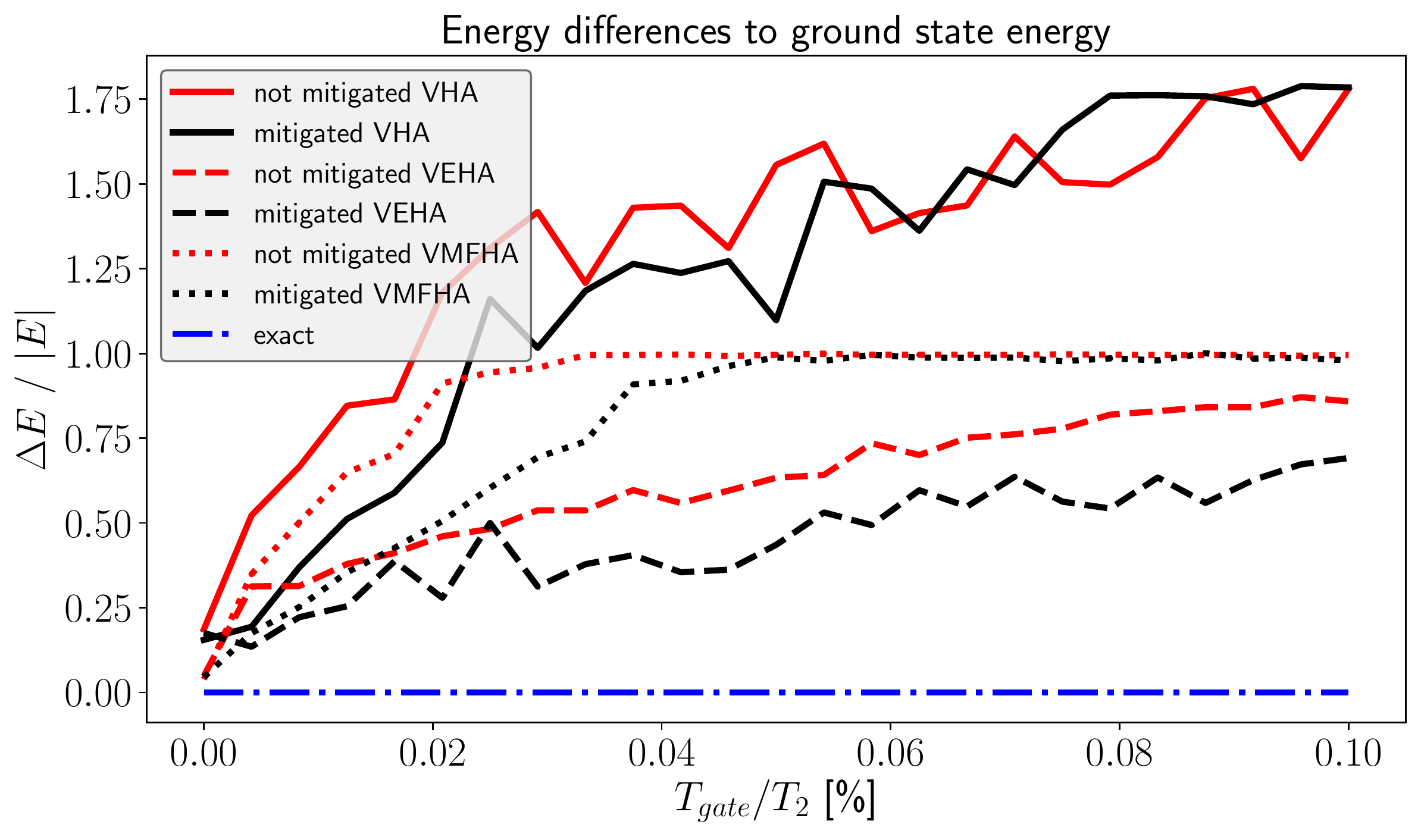}
  \caption{%
      \label{fig:2_2_E_stochastic}
      The energy difference between the calculated and the exact ground state energy scaled to the absolute value of the exact ground state energy for unmittigated (red) and mitigated (black) dephasing, for the VHA with post-selection (solid line), the VEHA (dashed line) and the VMFHA (dotted line) as a function of gate time over dephasing time at \(U = -3 \). Gate-based measurments were simulated with \(25000\) projective measurements. Compared to Fig.\ref{fig:2_2_E} we see the same overall trend but strong fluctuations in the results.
  }%
\end{figure}

\begin{figure}
  \centering
  \includegraphics[width=.5\textwidth]{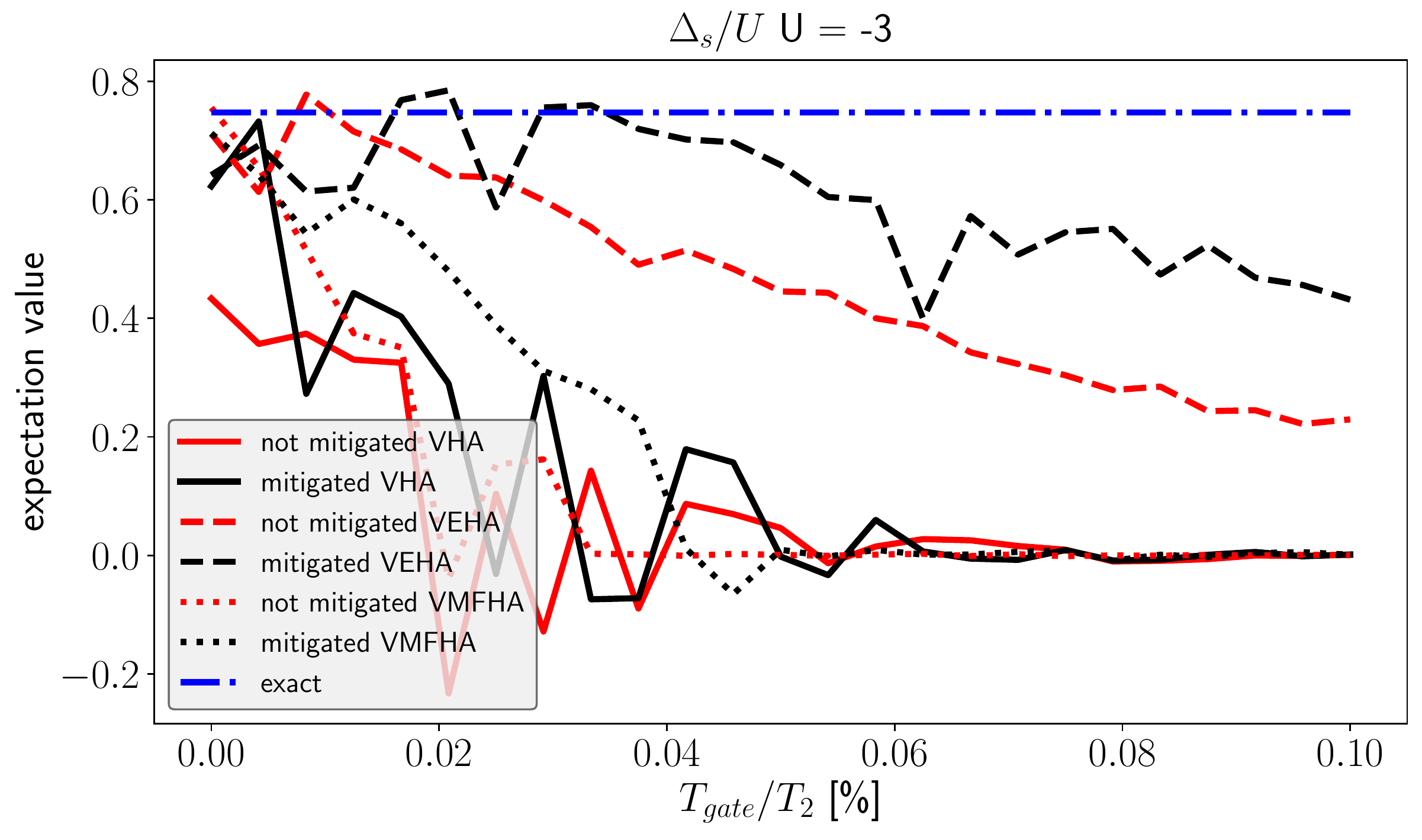}
  \caption{%
      \label{fig:2_2_exp_val_D_stochastic}
      The superconducting gap \(\Delta_s\) for exact diagonalization (blue dotted-dashed line) and for unmittigated (red) and mitigated (black) dephasing, for the VHA with post-selection (solid line), the VEHA (dashed line) and the VMFHA (dotted line) as a function of gate time over dephasing time at \(U=-3\). Gate-based measurements were simulated with \(25000\) projective measurements. Compared to Fig.\ref{fig:2_2_exp_val_D} we see the same overall trend but strong fluctuations in the results.
  }%
\end{figure}

The necessarily finite number of measurements leads to stochastic fluctuations in the measurement results. We show the effect for \(25000\) projective measurements per measured Pauli product and without assuming any measuring errors in Fig.\ref{fig:2_2_E_stochastic} and Fig.\ref{fig:2_2_exp_val_D_stochastic}. Here the stochastic fluctuations are too strong for meaningful predictions and the results are much worse compared to Fig.\ref{fig:2_2_E} and Fig.\ref{fig:2_2_exp_val_D}.
The number of projective measurements would need to be increased when running a calculation on an actual quantum computer.
\label{sec:stochastic}

\end{document}